\documentclass[aps,prd,twocolumn,superscriptaddress,
showpacs,preprintnumbers,nofootinbib,notitlepage]{revtex4-1}
\usepackage{amsmath,amssymb,hyperref,subfigure,xspace}
\usepackage{mciteplus,color,mathtools,graphicx}
\usepackage{bm,bbm,slashed}
\usepackage{multirow}
\usepackage[english]{babel}
\usepackage{mathrsfs}
\usepackage{enumerate}
\usepackage{xspace}
\usepackage{slashed}
\usepackage{wrapfig}
\usepackage[export]{adjustbox}
\usepackage[utf8]{inputenc}

\newcommand{\bcol}{\left[ \begin{array}{c}}
\newcommand{\ecol}{\end{array} \right]}
\newcommand{\beq}{\begin{equation}}
\newcommand{\eeq}{\end{equation}}
\newcommand{\beqq}{\begin{eqnarray}}
\newcommand{\eeqq}{\end{eqnarray}}

\newcommand{\bra}{\left\langle}
\newcommand{\ket}{\right\rangle}
\newcommand{\p}{\bm{p}}

\hypersetup{ 
    pdfnewwindow=true,      
    colorlinks=true,       
    linkcolor=blue,          
    citecolor=blue,        
    filecolor=blue,      
    urlcolor=blue        
} 

\begin{document}


\newcommand{\ceem}{Center for  Exploration  of  Energy  and  Matter,  
Indiana  University,  
Bloomington,  IN  47403,  USA}
\newcommand{\indiana}{Physics  Department,  
Indiana  University,  
Bloomington,  IN  47405,  USA}
\newcommand{\jlab}{Theory Center,
Thomas  Jefferson  National  Accelerator  Facility, 
Newport  News,  VA  23606,  USA}


\title{The Coulomb flux tube revisited}


\author{Sebastian~M.~Dawid}
\email[email: ]{sdawid@iu.edu}
\affiliation{\indiana}\affiliation{\ceem}

\author{Adam~P.~Szczepaniak}
\affiliation{\indiana}\affiliation{\ceem}\affiliation{\jlab}

\preprint{JLAB-THY-19-3018}


\begin{abstract}

We perform  $SU(2)$ Yang-Mills lattice simulation of the  electric  field distribution in the Coulomb gauge for different values of $\beta$ to further investigate the nature of the Coulomb flux tube.
\end{abstract}

\date{\today}
\maketitle


\section{Introduction}
\label{sec:intro}

In the $SU(N)$ Yang-Mills theory in the Coulomb gauge, there is an instantaneous interaction between color charges, which is analogous to the Coulomb force between electric charges. The key difference, however, is that unlike QED where the Coulomb energy is a function of the distance between sources, in the non-abelian case it is a functional of the gauge field. Therefore in order to obtain the Coulomb potential, it is necessary to specify the state of the gluons. The Coulomb potential is conventionally referred to the state in which external, static sources are suddenly added to a vacuum. In the following, we refer to this as the bare state. This is different from the adiabatic situation, when gluons have  time to respond to the presence of the external charges resulting in the true QCD eigenstate. We refer to the later as the minimal energy or Wilson state, since its energy is related to the expectation value of the large Wilson loop.

The Coulomb potential, $V_{\text{C}}(r)$, evaluated on a state containing a static quark-antiquark pair is of special interest. In the past few years, it has been extensively studied \citep{Cucchieri:2000, Cucchieri:2001, Szczepaniak:2001, Zwanziger:2002, Zwanziger:2003, Greensite:2003, Cucchieri:2003, Bowman:2004, Nakagawa:2006, Leder:2011, Golterman:2012, Greensite:2014, Reinhardt:2017, Cooper:2018} both within continuum and lattice approaches, which  contributed to a better understanding of the quark confinement \citep{Gribov:1977, Zwanziger:1998, Greensite:2011}. In particular, it has been shown that the Coulomb confinement is  necessary  for the Wilson confinement \citep{Zwanziger:2002}. This  has been confirmed on the lattice  \citep{Greensite:2003, Cucchieri:2003} where it was found that the Coulomb potential rises linearly for large quark-antiquark separation $r$, with  the associated string tension $\sigma_{\text{C}}$  larger by approximately a factor of three compared to  the minimal one, $\sigma$,  obtained from the expectation value of the large Wilson loop.

Since both potentials are confining, it is reasonable to ask how other gauge-field-related observables compare in the two states. Numerical studies with $SU(2)$ and $SU(3)$ lattice Yang-Mills theories have established a picture of a flux tube formation between a quark and an antiquark in the minimal energy state \citep{Fukugita:1983, Flower:1985, Wosiek:1987, Bronzan:1987, DiGiacomo:1990, Trottier:1993, Bali:1994, Cardaci:2010, Cea:2012, Cardoso:2013, Lukashov:2017, Bicudo:2018}. Phenomenologically, it was established that both the action and energy densities vanish exponentially in a direction perpendicular to the line joining the quark sources. In principle, these observables can be obtained by measuring a (normalized) correlation function between a large Wilson loop $W(r,t)$ and appropriately placed plaquette $U_P$, which serves as a chromo-electric (or chromo-magnetic) field probe \citep{Cardoso:2013}. 

The analogous question concerning the bare $Q{\bar Q}$ state was recently addressed by Chung and Greensite in Ref. \citep{Chung:2017}. There it was found that also the bare state has the flux-tube-like characteristics with an exponentially decaying transverse profile. This is an interesting and unexpected result since  analysis of the Coulomb energy density distribution and related observables, e.g. the ghost propagator, in the infinite volume \citep{Szczepaniak:2001, Bowman:2004, Watson:2010, Leder:2010} typically predicts a power-law fall-off. Also  one could  argue that the Gribov-Zwanzinger confinement proposal \citep{Gribov:1977, Zwanziger:1998}
implies long-range Van der Waals forces and thus the absence of flux tubes \citep{Bowman:2004}.

The aim of this paper is to shed more light on this result. Specifically, we extend the calculations of \citep{Chung:2017} that were performed at $\beta=2.5$ to a considerably better precision for data at larger transverse distances, $y$ away from the $Q\bar Q$ axis, and we also performed the calculation for $\beta=2.3$ and $2.7$. Additionally, to understand the flux tube development, we investigated the Euclidean time evolution of the energy density profile. Finally, we performed an analysis of the data using both power-law and exponential profiles. 

The paper is organized in the following way. In Sec. \ref{sec:lattice} we give a summary of the lattice setup and describe the measured observables. In Sec. \ref{subsec:res-coupl} we present the key results of our simulations for different $\beta$'s, and discuss the analytic models. Results of the Euclidean time evolution are presented in Sec. \ref{subsec:time-dep} followed by summary and conclusions in Sec. \ref{sec:con}. The complete data set is given  in App. \ref{app:A}.


\section{ Electric field distribution in the presence of static quarks }
\label{sec:lattice}

In Ref. \citep{Greensite:2003} it was shown that on the lattice in the Coulomb gauge, both the Coulomb and Wilson energies can be calculated from the expectation value of two Wilson lines:
\beqq
\label{eq:time-dep-potential}
a V(r,t) &=& \log \frac{\left\langle \text{Tr} [ L_t(\bm{0}) L_t^\dag(\bm{r}) ] \right\rangle}{\left\langle \text{Tr} [ L_{t+a}(\bm{0}) L_{t+a}^\dag(\bm{r}) ] \right\rangle} \ ,
\eeqq
where $a$ is the lattice spacing, and $L_t(\bm{x})$ is a time-like Wilson line of length $t$ starting at position $(0,\bm{x})$. In the (Euclidean time) limit $t\to \infty$, potential $V(r,t)$ becomes the Wilson eigenenergy $V_\text{min}(r)$. In the limit $t\to 0$ this quantity, up to an additive, $r$-independent constant, approaches the lattice version of $V_{\text{C}}(r)$, defined as a correlation of short time-like links
\beqq
\label{eq:lattice-coulomb}
a V(r, 0) &=& - \log  \left\langle \frac{1}{N} \text{Tr} [ U_0(0,\bm{0}) U_0^\dag(0,\bm{r}) ] \right\rangle \ ,
\eeqq
where $U_\mu(x)$ is a link variable  at position $x=(x^0,\bm{x})$ in the direction of $\mu \in [0,3]$. Four-vectors $(0,\bm{0})$ and $(0,\bm{r})$ represent positions of the quark and the antiquark, respectively. One can understand both limits as a starting and ending point of an equilibration process, which takes a set of gauge fields unperturbed by the presence of the $Q\bar Q$ pair and thermalizes it to the true ground state of the theory. As mentioned earlier, in the $SU(2)$ Yang-Mills theory, it was found that at fixed value of $\beta$, the string tension $\sigma_C$ computed from $V(r,t)$ as a function of $r$ decreases with increasing $t$ and approaches that of the Wilson energy at large times. In the Coulomb limit $t\to0$, it is found \citep{Greensite:2003} to be larger by approximately a factor of three, see Fig. \ref{fig:tension}.

\begin{figure}[b]
\begin{center}
\includegraphics[scale=0.6]{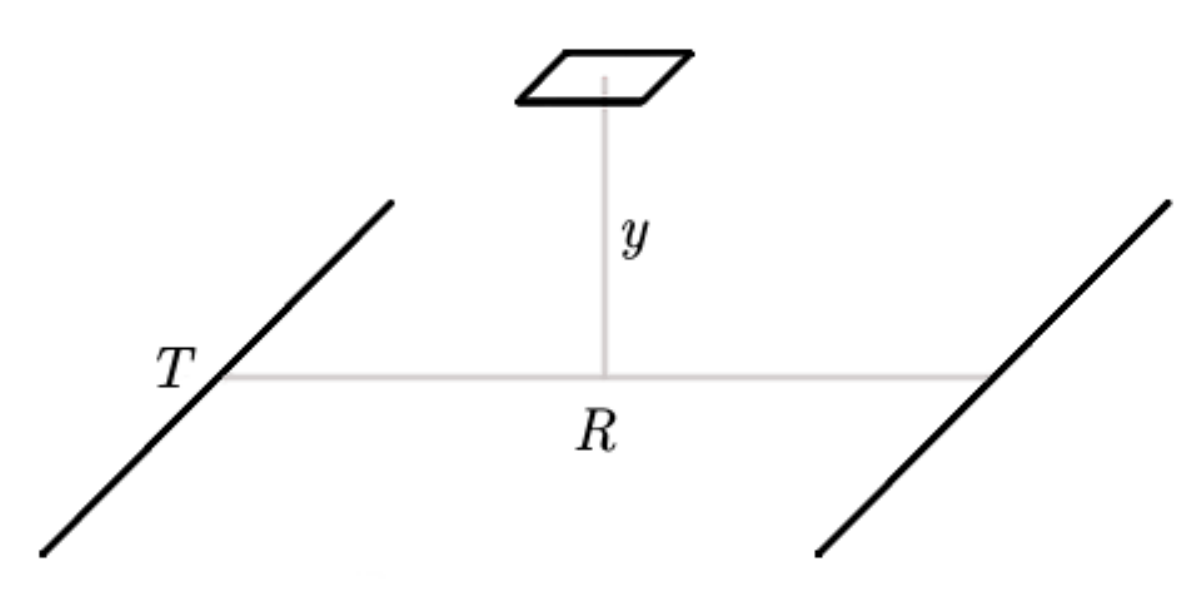}
\caption{Schematic picture of the arrangement of links which, after normalization, corresponds to the observable $Q_T(R,y)$ defined in Eq. \eqref{eq:Q}. The minimal energy flux tube is measured for large $T$'s with two Wilson lines $L_T(\bm{0})$ and $L_T^\dag(\bm{R})$ replaced by the Wilson loop $W(R,T)$ of size $R \times T$. \citep{Cardoso:2013} }
\label{fig:minimal}
\end{center}
\end{figure}

In the Coulomb gauge, the longitudinal component of the chromo-electric field $\bm{E}_{\text{L}}$ is determined by charge distribution (as in the classic theory) via the Gauss' law. Thus one can determine distribution of this field in a state at any time, as the expectation value of $\bra \text{Tr} \bm{E}^2_{\text{L}} \ket$. Since it is expected that the main contribution to the energy density comes from the field component parallel to the $Q\bar{Q}$ axis, which we assume to lie in the $x$ direction, here we calculate this one component contribution as in Ref. \citep{Chung:2017} using
\beq
\label{eq:Q}
Q_T(R,y) \!=\! \frac{  \bra \text{Tr} [ L_T(\bm{0}) L_T^\dag(\bm{R})] \frac{1}{2} \text{Tr} U_\text{P}(\p,T)\ket }{\bra \text{Tr} [ L_T(\bm{0}) L_T^\dag(\bm{R})] \ket} 
- \frac{1}{2} \bra \text{Tr} U_\text{P} \ket \ ,
\eeq
where we have switched to dimensionless distances $R=r/a$ and $T=t/a$. Plaquette $U_{\text{P}}(\bm{p},T)$ is defined analogously to $U_{\text{P}}(\bm{p},0)$ in Eqs. (25) and (26) of Ref. \citep{Chung:2017}, i.e. it is oriented in $xt$-plane and placed at position
\beq
\label{eq:p}
\bm{p} = \left\{ \begin{array}{cc}
\frac{R}{2} \bm{\hat{e}_x} + y \bm{\hat{e}_y } ,  & \text{for even }R \ , \vspace{5pt} ~ \\ 
\frac{R-1}{2} \bm{\hat{e}_x}  + y \bm{\hat{e}_y } , &  \text{for odd }R \ ,
\end{array} \right.
\eeq
but at a different time slice: $T/2$ for even $T$, or $(T-1)/2$ for odd $T$. Here $y=y_\perp/a$ is a distance from the $Q\bar{Q}$ axis. The plaquette acts as a probe of the $x$-component of the longitudinal chromo-electric field at point $\bm{p}$, therefore allows to see how $Q_T$ changes with the transverse distance $y$. The key quantity, which is the numerator of the first term in Eq. \eqref{eq:Q}, is depicted schematically in Fig. \ref{fig:minimal}.

The observable $Q_T(R,y)$ is a generalized version of $Q(R,y)$ introduced in Eq. (24) of Ref. \citep{Chung:2017}, and reduces to the latter for $T=1$, i.e. when the Wilson line is equal to one temporal link $L_{T=1}(\bm{x}) = U_0(0,\bm{x})$. From $Q(R,y)$, in Ref. \citep{Chung:2017}, the energy distribution in the Coulomb state was obtained. The generalization given by Eq. \eqref{eq:Q} allows us to make a connection with the minimal energy flux tube measured as a normalized correlation function between a large Wilson loop and a plaquette \citep{Cardoso:2013}. Since in the Coulomb gauge spatial links become very close to the identity matrix, the Wilson loop can be approximated as a product of two temporal Wilson lines, and thus one should be able to observe the convergence of the bare state field distribution to the Wilson state field distribution as increasing $T$ is considered.


\subsection{Lattice framework}
\label{subsec:lattice}

We performed Monte Carlo simulation of the pure $SU(2)$ Yang-Mills theory using Wilson's action \citep{Wilson:1974}. We used lattice of size $V=32^4$ with periodic boundary conditions, at couplings $\beta = 2.3, \hspace{2pt} 2.5, \hspace{2pt} 2.7$, which correspond to lattice spacings $a = 0.165, \hspace{2pt} 0.085,\hspace{2pt} 0.045$ fm, respectively \citep{Bloch:2004}. Field configurations were generated using the heat-bath algorithm \citep{Creutz:1980} and we considered the lattice equilibrated after initial $10 \hspace{2pt} 000$ sweeps for each $\beta$. Lattice configurations used for data extraction were separated by $300$ sweeps to minimize the impact of autocorrelations. The Coulomb gauge was assumed to be fixed when $\Delta F = |F_i-F_{i+1}| < 10^{-7}$, where
\beq
F_i = \frac{1}{4V} \sum_{\mu=1}^3 \sum_x \text{Tr} \hspace{2pt} U_\mu(x) \ 
\eeq
is a value of the functional to be minimized, after the $i$-th gauge fixing iteration. We found that the more strict condition $\Delta F < 10^{-8}$ did not affect the results noticeably. To speed up the gauge fixing procedure we implemented an overrelaxation method \citep{Mandula:1990,Giusti:2001} with $\omega=1.75$.

As a check, we repeated the simulation of Ref. \citep{Chung:2017} with increased  statistics  by generating $30\hspace{2pt}000$ lattice configurations at $\beta=2.5$, to obtain $Q_{T=1}(R,y)$. In addition we used approximately $11\hspace{2pt}000$ configurations to compute $Q$ for $\beta=2.3$ and $\beta=2.7$. The Euclidean time dependence was obtained also from the same number of configurations at $\beta=2.5$ and for $T=1,2,3,4,5$. For each configuration, we averaged the value of the observables over four possible translations and three $90^\circ$ spatial rotations. Expectation values of observables and statistical errors were obtained from the jackknife method. We do not investigate here the systematic errors due to e.g. finite lattice size or gauge fixing quality and assume they do not affect the main conclusion of our work.


\section{Results}
\label{sec:results}

\begin{figure}[b]
\subfigure[~R=2]
{
\includegraphics[width=0.4\textwidth]{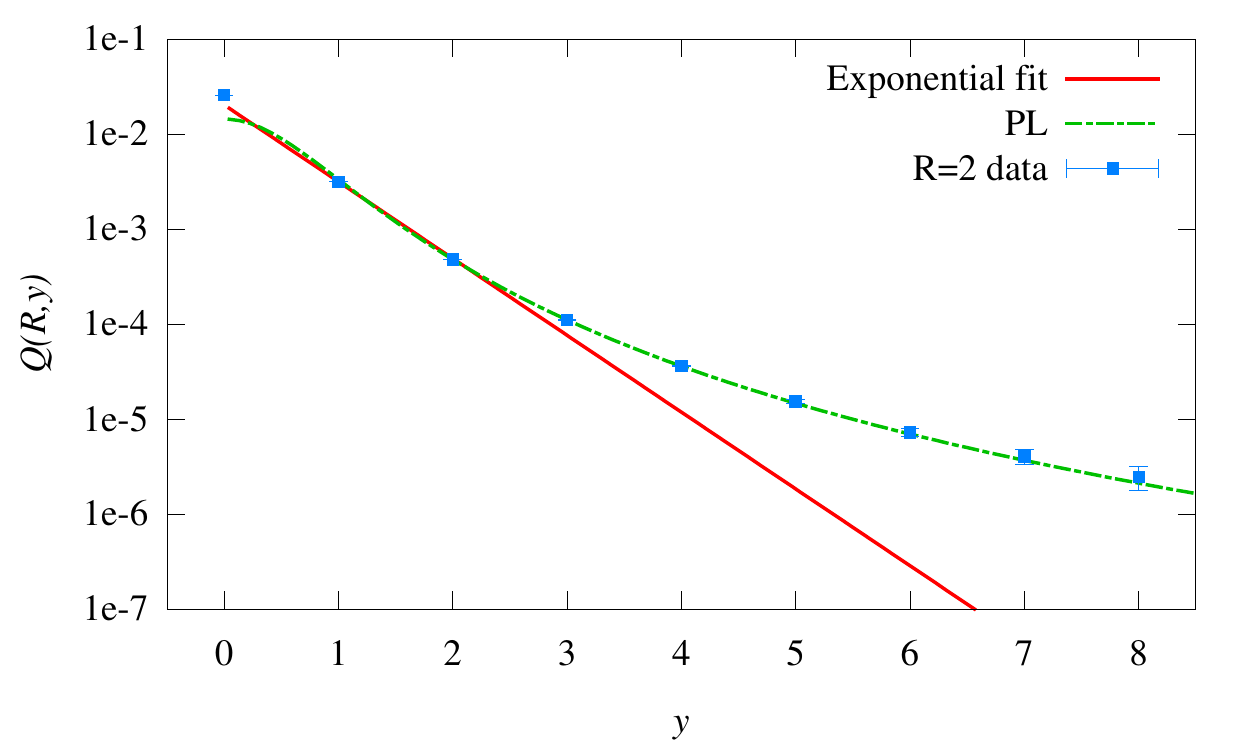}
\label{fig:5R2}
}
\subfigure[~R=7]
{
\includegraphics[width=0.4\textwidth]{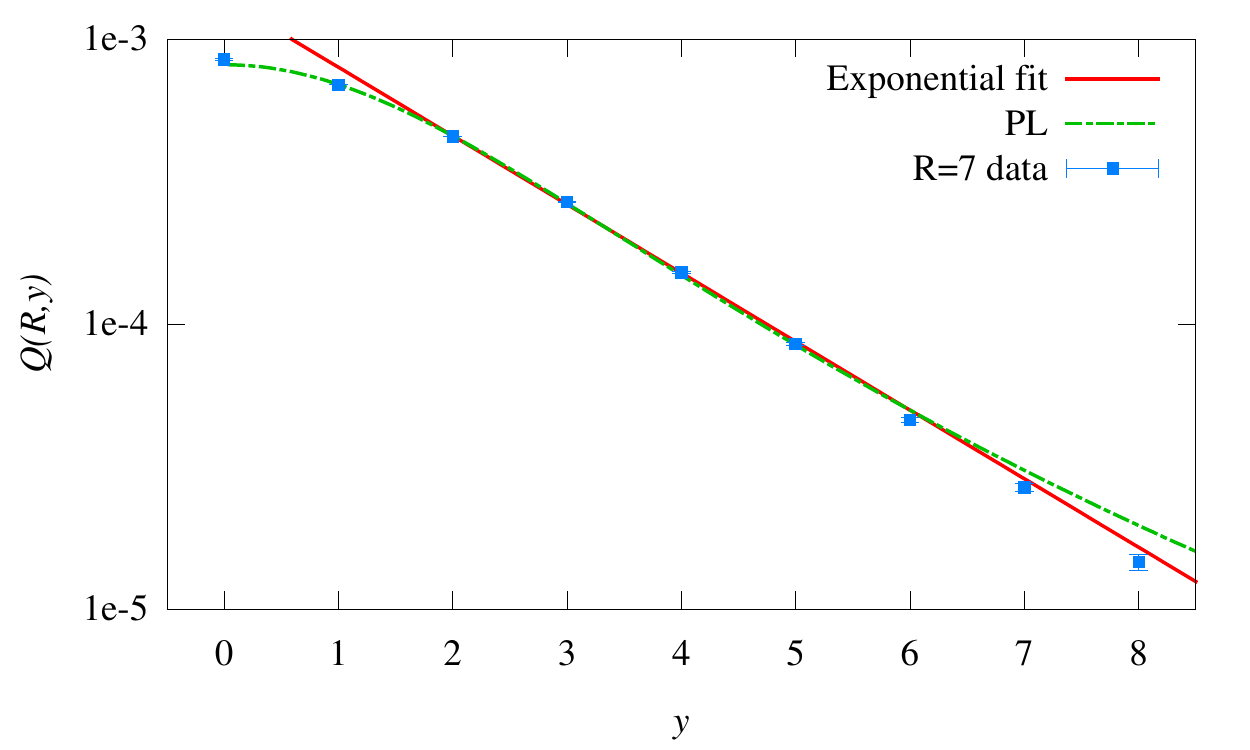}
\label{fig:5R6}
}
\caption{Dependence of $Q_{T=1}$ on transverse distance $y$ at fixed quark-antiquark separation $R$ (in lattice units). Red, solid line is an exponential fit (CG), and green, dashed line is a power-law (PL) fit, both described in Sec. \ref{subsec:res-coupl}. Data points were obtained from $30\hspace{2pt}000$ gauge-fixed lattice configurations at $\beta=2.5$, for lattice volume $V=32^4$.}
\label{fig:5ydep}
\end{figure}


\subsection{Results for different lattice couplings}
\label{subsec:res-coupl}

\begin{table}[b]
\begin{ruledtabular}
\begin{tabular}{ccccc}
$R$ & $A$ & $B$ & fit interval & $\chi^2/$d.o.f \\  \hline
1 & 3.20(6) & 2.45(6) & [1,3] & 779 \\ 
2 & 3.89(7) & 1.89(7)  & [1,3]  & 2586 \\
3 & 4.81(3) & 1.29(4) & [1,4] & 1038  \\
4 & 5.40(3) & 1.01(2) & [1,6] & 394 \\
5 & 6.046(8) & 0.732(5) & [1,7] & 28 \\
6 & 6.298(7) & 0.643(3) & [2,7] & 1.5 \\
7 & 6.58(2) & 0.554(8) & [2,7] & 15 \\
8 & 6.79(3) & 0.50(1) & [2,7]  & 27 
\end{tabular}
\end{ruledtabular}
\caption{The CG fit parameters, for different $R$'s at $\beta=2.5$. We followed the fitting procedure from Ref. \citep{Chung:2017} and deviated from it only for $R=6,7,8$, for which fit intervals are $[2,7]$ compared to $[1,7]$ in Ref. \citep{Chung:2017}. For $R\leqslant5$ our values agree with Tab. I  in Ref. \citep{Chung:2017}.
}
\label{tab:CG25}
\end{table}

\begin{table}[t]
\begin{ruledtabular}
\begin{tabular}{ccccc}
$R$ & $a$ & $b$ & fit interval & $\chi^2/$d.o.f \\
\hline
1 & 0.008(3)  & 2.20(9)  & [3,8]  &  1.41  \\ 
2 & 0.00421(6) & 2.114(4) & [2,8]  & 0.33  \\
3 & 0.00255(7) & 1.996(8) & [2,8] & 2.52 \\
4 & 0.0025(3) & 2.01(3) & [3,8] &  4.40 \\
5 & 0.0026(4) & 2.02(4) & [3,8] & 10 \\
6 & 0.0025(2) & 2.01(2) & [2,8] & 13 \\
7 & 0.0029(4) & 2.04(3) & [2,8] & 19 \\
8 & 0.0044(7) & 2.12(4) & [2,8] & 15
\end{tabular}
\end{ruledtabular}
\caption{PL fit parameters, for different $R$'s at $\beta=2.5$. It is worth noticing that the power $b \approx 2$ is in agreement with a prediction for the large $y$ behavior of the flux tube transverse profile from Ref. \citep{Bowman:2004}. Using larger fit intervals usually increases the value of $b$ to $b \approx 2.3$-$2.7$.}
\label{tab:PL25}
\end{table}

In the case of the Wilson state, it was shown in  \citep{Bali:1994} that the profile of the Wilson flux tube, for small separations $R$, can be calculated from  perturbation theory\footnote{See Eq. (33) in Sec. III of Ref. \citep{Bali:1994}.} and to the leading order in $\alpha_s$, falls off as $1/y^6$. For large quark separations it is observed that the Wilson energy density profile changes from power-law to exponential \citep{Cardoso:2013}. In this subsection we discuss the energy distribution in the Coulomb state,   
$Q_{T=1}(R,y)$ calculated at different values of the coupling $\beta$. To understand dependence on the transverse distance $y$, we first try two simple models, one is the power law, (PL) $Q_{\text{PL}}$, motivated by \citep{Bali:1994} and the other exponential (CG) $Q_{\text{CG}}$ used in Ref. \citep{Chung:2017}  
\beqq
\label{eq:PL}
Q_{\text{PL}}(R,y) &=& \frac{16 a R^2}{(R^2+4y^2)^b} \ , \\
\label{eq:CG}
Q_{\text{CG}}(R,y) &=& \exp\left(-A-By\right) \ . 
\eeqq
While $b=3$  is predicted for the Wilson state, analytical calculations in the Coulomb gauge predict $b \approx 2$, independent of $R$ \citep{Bowman:2004}. Thus in the Coulomb gauge one might expect power-law behavior in $y$ for small values of $R$ and as a consequence, it is the $y$-behavior of the energy profile at large $R$ that should be examined to discriminate between an exponential and power-law decay.

Our most precise measurement was performed at  $\beta=2.5$. Sample  plots of the energy transverse profile for (``small") $R=2$ and (``large") $R=7$ are shown in Fig. \ref{fig:5ydep}, and for all other quark separations are summarized in  Fig. \ref{fig:allR-25} in App. \ref{app:A}. The fit parameters, fit intervals and corresponding values of $\chi^2/$d.o.f. for CG and PL models are presented in Tabs. \ref{tab:CG25} and \ref{tab:PL25}. Details of the fitting procedure are given in the App. \ref{app:A}. It appears that the data favor the PL model over the simple exponential (CG), however, it does so with varying quality depending on $R$. Specifically, for large $R$, we could not reach conclusive results and both models appear to be deficient. This may indicate that at $\beta=2.5$, $Q_{T=1}(R,y)$ does not properly reproduce the Coulomb energy density distribution. Since the Coulomb energy is obtained in the continuum limit, or at least for large values of $\beta$, it would appear that $\beta=2.5$ is not large enough. 

\begin{figure}[t]
\subfigure[~$r\approx0.34$ fm]
{
\includegraphics[width=0.4\textwidth]{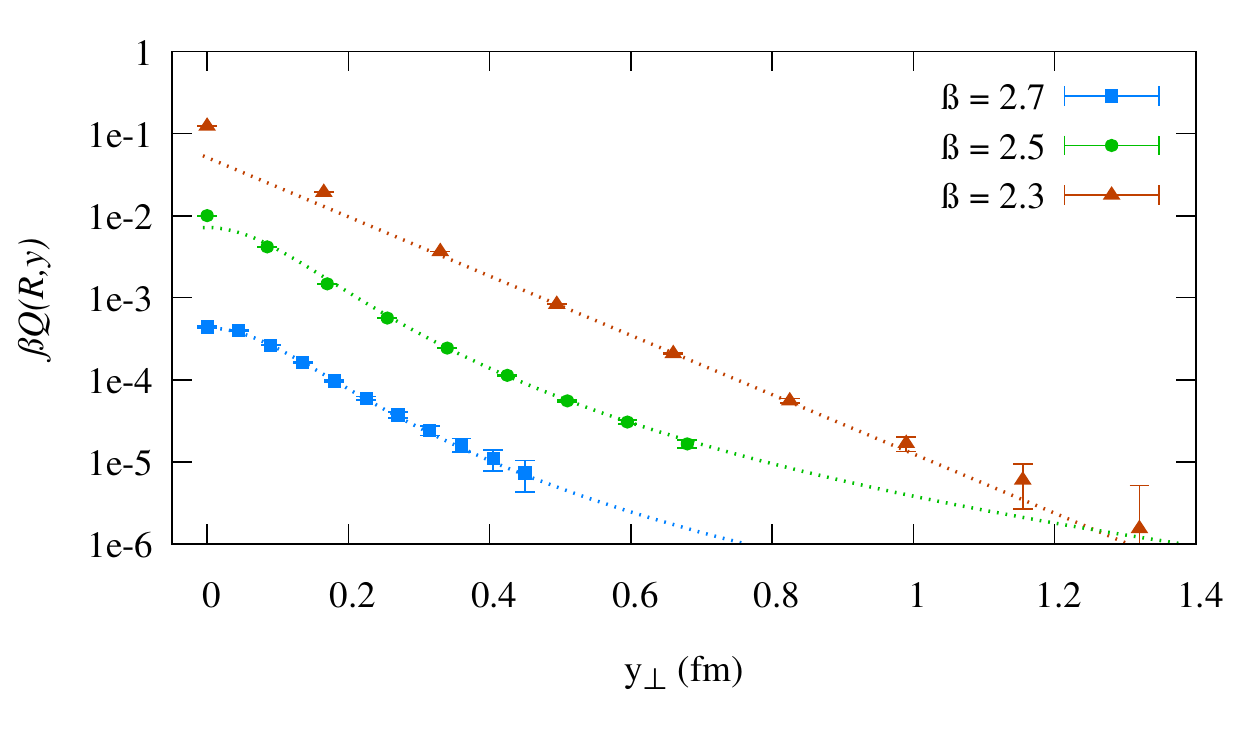}
\label{fig:betas-a}
}
\subfigure[~$r\approx0.51$ fm]
{
\includegraphics[width=0.4\textwidth]{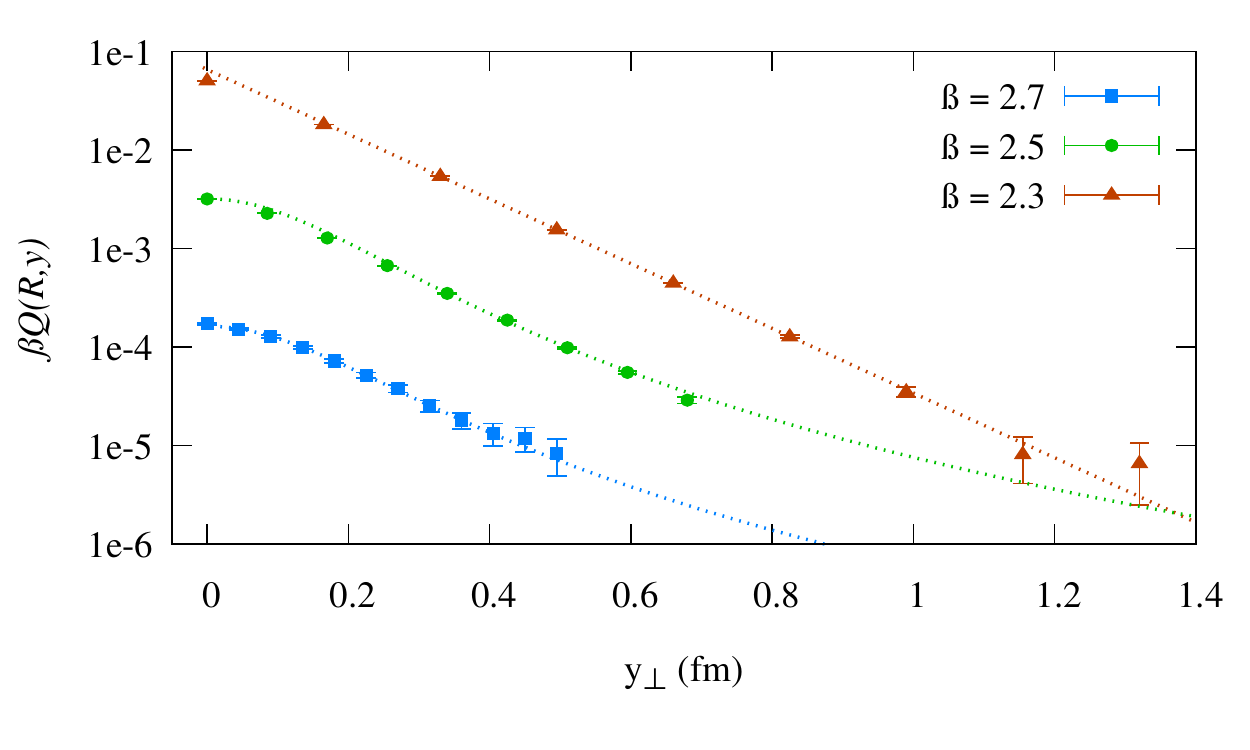}
\label{fig:betas-b}
}
\caption{Dependence of $\beta Q_{T=1}$ on transverse distance $y_\perp = a y$ at fixed quark-antiquark separation $r=aR$. Fig. \ref{fig:betas-a} shows result for physical separations $r \approx 0.34$ fm and Fig. \ref{fig:betas-b} for $r \approx 0.51$ fm. The dotted lines represent the best fits: CG for $\beta=2.3$, PL for $\beta=2.5$ and for $\beta=2.7$. The energy density shape for $\beta=2.3$ shows a different behavior from the profiles at larger values of the coupling, even for relatively small values of $R$.}
\label{fig:betas-compared}
\end{figure}

We thus hypothesize that $Q_{T=1}(R,y)$ evolves from a power-law behavior for large $\beta$ to an exponential (small $\beta$) and consequently performed a calculation at $\beta=2.3$ and $\beta=2.7$. These values correspond to approximately constant ratios of the lattice spacing $a_{\beta=2.3}/a_{\beta=2.5} \approx a_{\beta=2.5}/a_{\beta=2.7} \approx 2$. For example, the energy distribution profile at $\beta=2.5$ for $R=2$ can be compared with that at $\beta=2.3$ for $R=1$, and at $\beta=2.7$ for $R=4$, and all of them correspond to $r=aR \approx 0.17$ fm. In Fig. \ref{fig:betas-compared} we show the energy profiles for the three values of $\beta$. One can see that indeed for larger $\beta$'s the Coulomb flux tube profile appears to follow a power-law, while at the lowest value, $\beta=2.3$, already starting at low values of $R$, $R=2$, it appears much closer to an exponential -- as discussed above for small $R$ the $y$-profile is expected to follow a power-law. 

Taking a closer look at the $\beta=2.3$ data, shown in Fig. \ref{fig:allR-23} in App. \ref{app:A}, we see a clear exponential decay for large $R$. To see if the energy distribution for this value of the coupling $\beta$ can be described by the Wilson flux tube profile, we employed an ``improved" exponential model (CCB):
\beq
\label{eq:CCB}
Q_\text{CCB}
= \exp\left(-\frac{2}{\lambda} \sqrt{y^2+\nu^2} - 2 \frac{\nu}{\lambda} \right) \ ,
\eeq
proposed\footnote{Here we write $-2\nu/\lambda$ compared to their $+2\nu/\lambda$ in the exponential.} in Ref. \citep{Cardoso:2013}. Here $\nu$ and $\lambda$ are free parameters and their values for our data set can be found in Tab. \ref{tab:23CG}. The CG and CCB models are similar, as they both describe the exponential decay for large $y$. However CCB model also includes flattening of the energy profile close to the $Q\bar{Q}$ axis. From the plots in Fig. \ref{fig:allR-23} it can be seen that the fit accuracy is improved as $R$ increases. 
For $R=1$, were a power law is expected we fitted the $\beta=2.3$ data with PL model in an interval $y \in [3,8]$, obtaining  $b=2.83(7)$, which is close to the perturbative prediction of $b=3$ for the minimal state energy density shape. Even though it might look that the plots for $R=2$ and even $R=3$ might follow a power law, we were not able to obtain a good fit with such models.

\begin{table}[t]
\begin{ruledtabular}
\begin{tabular}{ccccc}
$R$ & $\nu$ & $\lambda$ & fit interval & $\chi^2/$d.o.f \\  \hline
1 & --- & ---  & ---  &  --- \\ 
2 & 1.73(3) & 1.31(1) & [3:8]  & 1.38  \\
3 & 1.77(1) & 1.438(6) & [3:8] & 0.77 \\
4 & 1.93(1) & 1.567(5) & [3:8] &  0.54 \\
5 & 2.11(2) & 1.699(8) & [3:8] & 1.24 \\
6 & 2.314(4) & 1.820(2) & [2:8] & 0.51 \\
7 & 2.587(4) & 1.973(3) & [1:8] & 0.84 \\
8 & 2.835(7) & 2.106(5) & [1:8] & 0.97 \\
\end{tabular}
\end{ruledtabular}
\caption{As explained in the text, the small-$\beta$ profile is expected to be well described by the CCB model, except for small $R$. The parameters for different $R$'s are shown. For $R=1$ the CCB model were not able to reproduce the data. In this case the PL fit was used in an interval $y \in [3,8]$, and with parameters $a=0.23(5)$, $b=2.83(7)$ yielding $\chi^2$/d.o.f.$=0.71$. }
\label{tab:23CG}
\end{table}

\begin{table}[b]
\begin{ruledtabular}
\begin{tabular}{ccccc}
$R$ & $a$ & $b$ & fit interval & $\chi^2/$d.o.f \\  \hline
1 & 0.02(4)  & 2.75(5)  & [2:8]  &  0.73  \\ 
2 & 0.0055(5) & 2.46(3) & [2:8]  & 0.94  \\
3 & 0.00305(3) & 2.330(4) & [1:8] & 0.57 \\
4 & 0.0022(3) & 2.28(3) & [2:8] &  2.05 \\
5 & 0.0013(1) & 2.16(2) & [2:8] & 1.30 \\
6 & 0.0013(2) & 2.17(4) & [2:8] & 2.09 \\
7 & 0.0008(1) & 2.07(4) & [2:8] & 0.97 \\
8 & 0.0009(1) & 2.09(3) & [2:8] & 0.55 \\
\end{tabular}
\end{ruledtabular}
\caption{Best PL fit parameters for different values of $R$ at $\beta=2.7$. Again $b \approx 2$ agrees with Ref. \citep{Bowman:2004}. The power depends significantly on an interval $[y_{\text{min}},8]$ used for fitting, and becomes closer to the asymptotic value $b=2$ for larger $y_{\text{min}}$. It signalizes that the simple PL fit describes the large $y$ behavior correctly but needs to be improved to incorporate the close-to-axis shape.}
\label{tab:PL27}
\end{table}

\begin{figure}[t]
\subfigure[~$R=2$]
{
\includegraphics[width=0.4\textwidth]{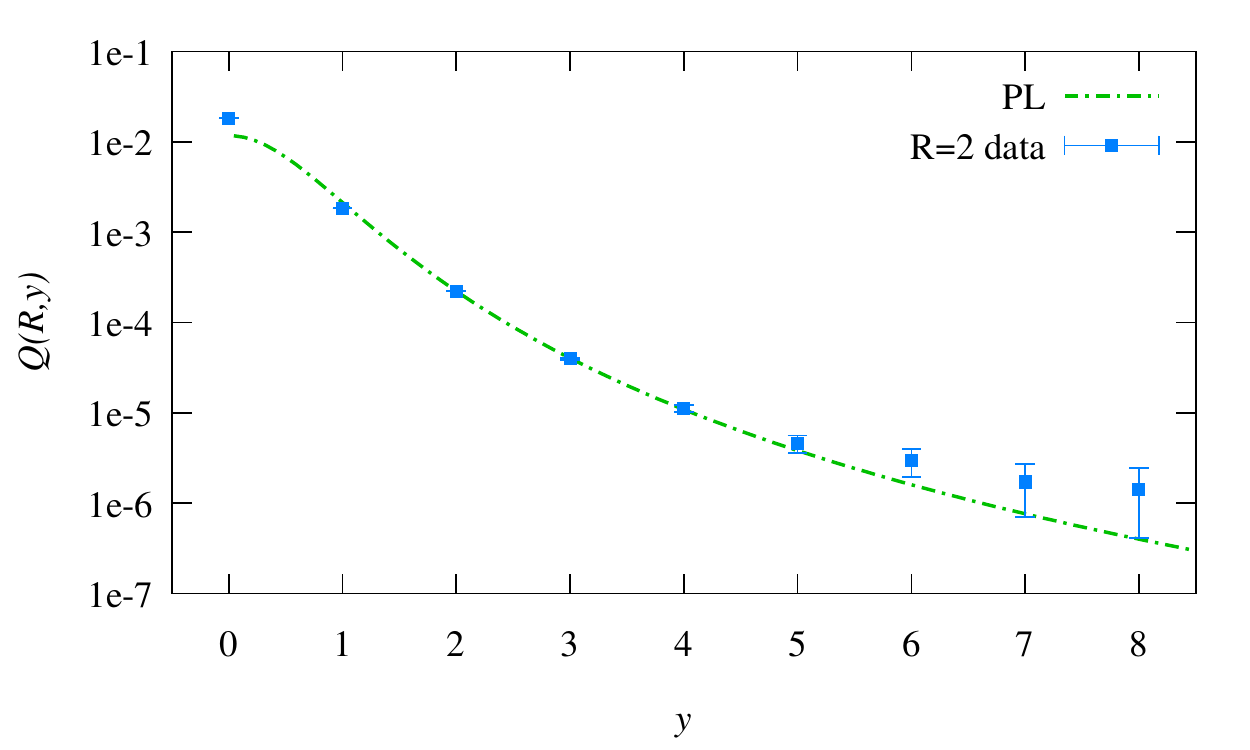}
\label{fig:7R2}
}
\subfigure[~$R=7$]
{
\includegraphics[width=0.4\textwidth]{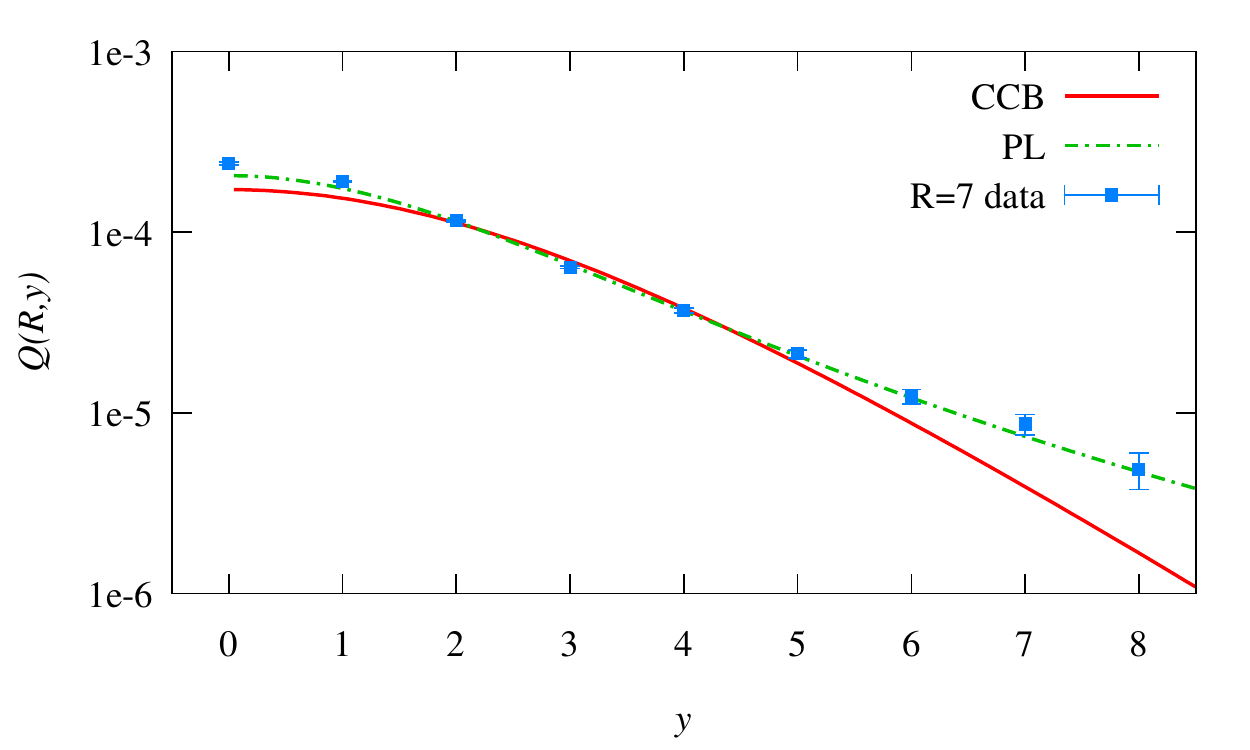}
\label{fig:7R7}
}
\caption{Dependence of $Q_{T=1}$ on transverse distance $y$ at fixed quark-antiquark separation $R$ (in lattice units). Green, dashed curve is the PL fit. The red one is the CCB fit included as an example how the exponential model fails to describe the data. Data points were obtained from $11\hspace{2pt}000$ gauge-fixed lattice configurations at $\beta=2.7$, for lattice volume $V=32^4$.}
\label{fig:7ydep}
\end{figure}

\begin{figure}[t]
\includegraphics[width=0.4\textwidth]{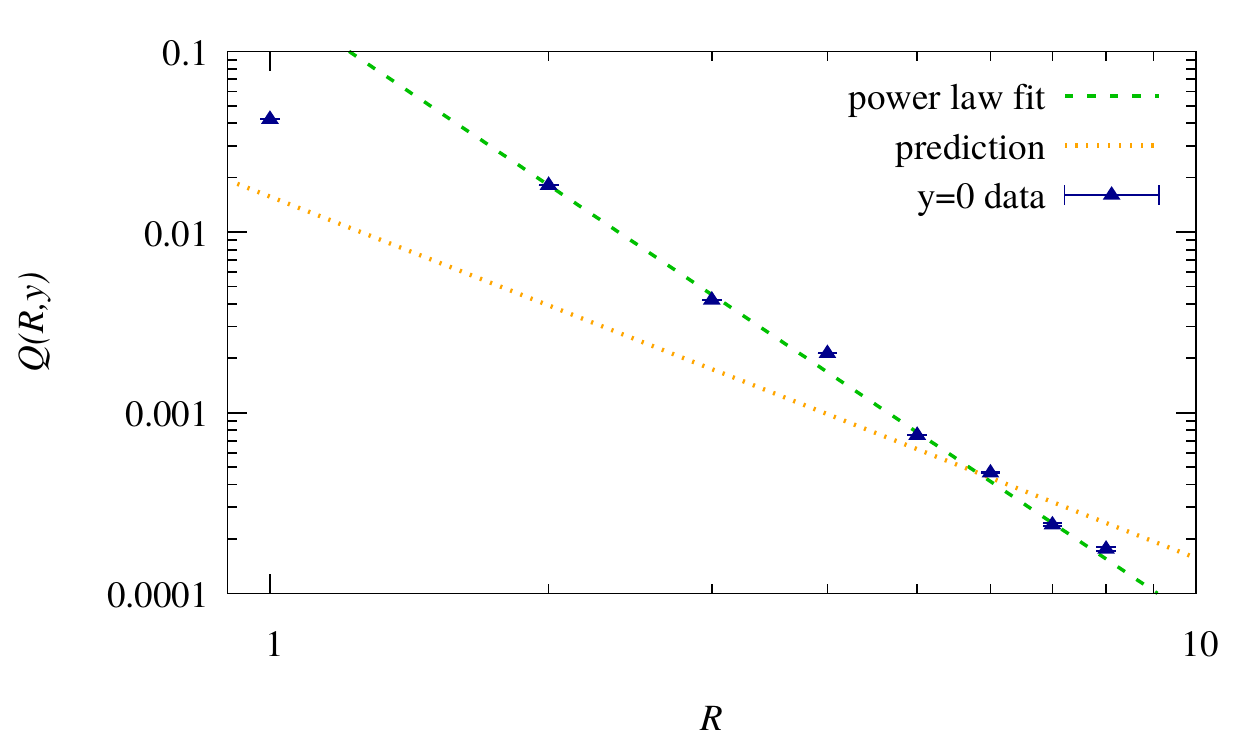}
\caption{The log-log plot of $Q_{T=1}$ at the middle point $y=0$ as a function of the quark-antiquark separation $R$. Green, dashed line is the $Q = a/R^b$ fit, with parameters $b = 3.4(1)$ and $a = 0.20(2)$. The $R=1$ point was excluded from the fit, because it does not satisfy $\mu R \gg 1$ condition from Ref. \citep{Bowman:2004} for $\mu=0.63$. The dotted, orange line represents theoretical prediction $\beta Q = 32 \sigma_\text{C}/ \pi^3 R^2$. The Coulomb string tension $\sigma_\text{C}$ at $\beta=2.7$ was obtained as a by-product of the energy density profile calculation from Eq. \eqref{eq:lattice-coulomb}. It was extracted from the fit $V_\text{C}(R) = \sigma_\text{C} R + \beta /R + \gamma$, with free parameters $ \sigma_\text{C}, \gamma, \beta$, and is equal to $\sigma_\text{C} = 0.0409(3)$.}
\label{fig:rdep0}
\end{figure}

For $\beta=2.7$ the energy density transverse profile  appears to follow a power law for all values of $R$ and neither CG nor CCB model gives a comparable description, see Fig. \ref{fig:7ydep} and Fig. \ref{fig:allR-27} in App. \ref{app:A}. The best PL fit parameters can be found in Tab. \ref{tab:PL27}. The fit indicates that the profile falls off approximately as $1/y^4$ with a transverse distance $y$, in agreement with Ref. \citep{Bowman:2004}. The exact power depends on an interval used for fitting, converging to $2$ when more of the low-$y$ points are excluded from the fit. The profile of the bare state energy density close to the $Q\bar{Q}$ axis could not be satisfactorily reconstructed by the straightforward PL model, which should be improved e.g. by inclusion of a factor correcting the small-$R$ and small-$y$ dependence. Such an factor should also mimic the change of the power $b\approx 2.7$ for small $R$ to $b\approx2$ for large $R$. We have tried to employ a perturbative prediction from Eq. (28) of Ref. \citep{Bowman:2004}
obtained from the Dyson-Schwinger equations in the Coulomb gauge. It did not lead to a conclusive results, predicting even more sever flattening of the small-$y$ shape at any value of $R$, than the PL model. In particular, it predicts a fall-off of the energy density at the middle point between quarks, $y=0$, as $Q \propto R^{-2}$, while our data set favors a higher power, see Fig. \ref{fig:rdep0}.


\subsection{The Euclidean time dependence}
\label{subsec:time-dep}

\begin{figure}[t]
\includegraphics[scale=0.6]{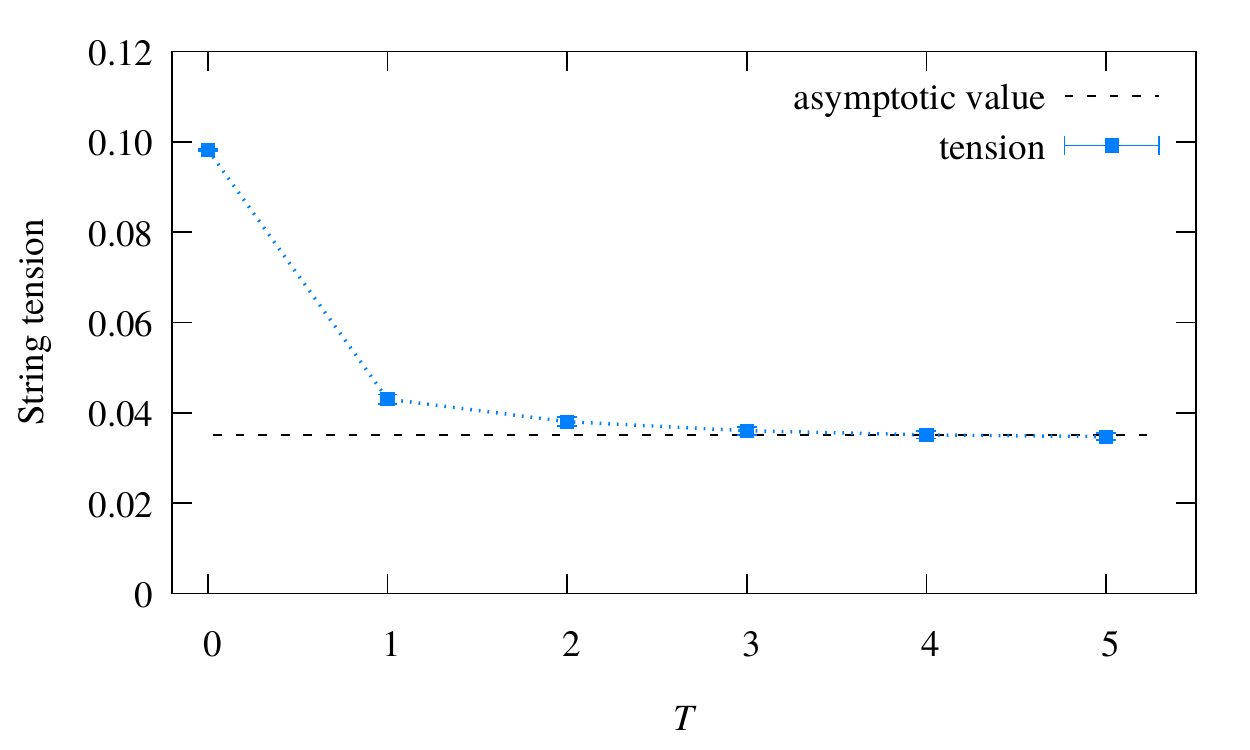}
\caption{Dependence of the string tension on $T$. It was obtained following the procedure of Ref. \citep{Greensite:2003}, for $\beta=2.5$ and around 1700 lattice configurations. The horizontal line is the asymptotic value from Ref. \citep{Bali:1994} } 
\label{fig:tension}
\end{figure}

As already discussed, the time development of the Coulomb potential and its evolution towards the minimal potential in principle can be analyzed by considering the large $t$ limit in Eq. \eqref{eq:time-dep-potential}. We extracted the behavior of the string tension as the Euclidean time progresses and observe its convergence to the minimal string tension, see Fig. \ref{fig:tension}, which agrees with Ref. \citep{Greensite:2003}.

\begin{figure}[t]
\subfigure[~$R=2$]
{
\includegraphics[width=0.4\textwidth]{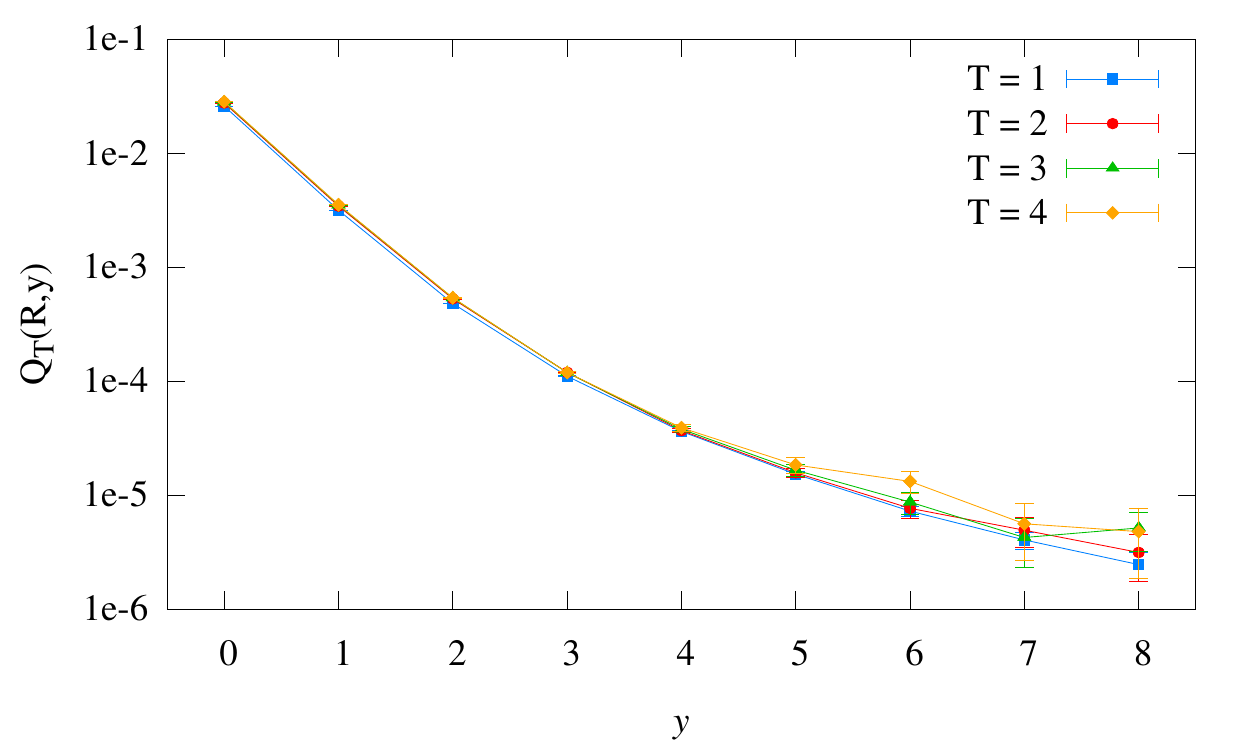}
\label{fig:2Rt}
}
\subfigure[~$R=7$]
{
\includegraphics[width=0.4\textwidth]{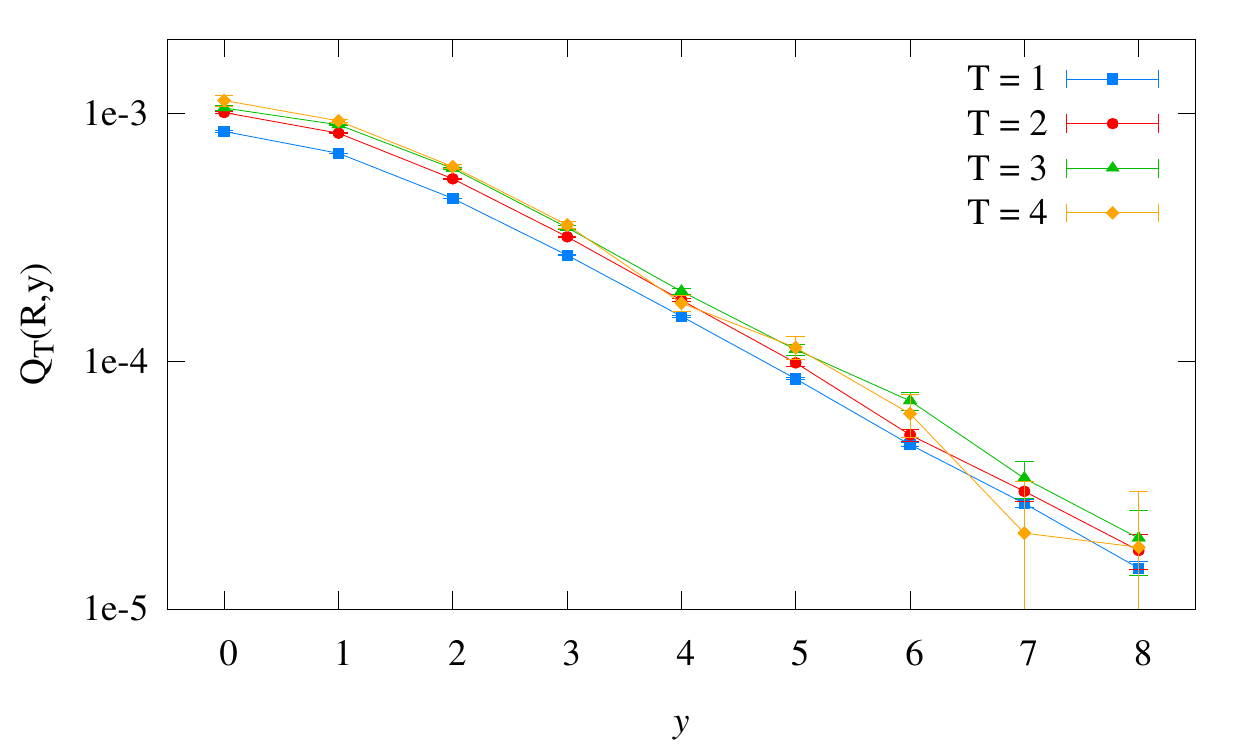}
\label{fig:7Rt}
}
\caption{Dependence of $Q_{T}(R,y)$ on transverse distance $y$ at fixed quark-antiquark separation $R$ for different $T$'s. Lines joining the points are included to guide the eye. Data points were obtained from around $11\hspace{2pt}000$ gauge-fixed lattice configurations at $\beta=2.5$, for lattice volume $V=32^4$. Result for $T=5$ is not presented for clarity of the plots. }
\label{fig:two-times}
\end{figure}

We have already seen that, as one takes relatively small value of $\beta$, it is possible to obtain an exponential fall-off of the bare state energy density transverse profile for large values of $R$. An interesting question arises if it converges to the minimal flux tube when the limit $T\to \infty$ is taken in Eq. \eqref{eq:Q} for larger values of $\beta$, for which a power-law behavior was found instead. As already discussed, in the Coulomb gauge the spatial links can be approximated by the identity matrix. If we thus made an assumption that the Wilson loop $W(R,T) \approx \text{Tr} \left[ L_T(\bm{0}) L_T^\dag(\bm{R}) \right]$, then for large values of $T$ Eq. \eqref{eq:Q} would agree with Eq. (7) from Ref. \citep{Cardoso:2013}, i.e. the formula for the chromo-electric contribution to the energy density profile of the minimal flux tube. To answer this question we investigated $Q_T(R,y)$ for $T$ from $T=1$ up to $T=5$, and the results are presented in Fig. \ref{fig:two-times} for $R=2$ and $R=7$, and Fig. \ref{fig:all-times} in App. \ref{app:A} for all $R$'s. From the plots it seems that the quantitative behavior of the profile does not change with the growing $T$; it increases globally (i.e. for all values of $y$) but the functional form of the profile appears unaltered. This is quite surprising. It might indicate that the flux tube described by $Q_{T=1}$ at $\beta=2.5$ is already ``equilibrated" in a sense that it represents closely the minimal flux tube rather than the bare state energy density distribution, and one should move closer towards the continuum limit (larger $\beta$) to see a noticeable difference. This requires further investigation, with a better statistics and in a larger volume, where larger values of $R$, $y$ and $T$ can be considered.

\section{Conclusions}
\label{sec:con}

In this paper we have extended the calculation of \citep{Chung:2017} to other $\beta$ values. We found that our numerical results for $Q_{T=1}$ at $\beta=2.5$ agree with those presented in Ref. \citep{Chung:2017} with increased statistics. The concussion from \citep{Chung:2017} was that the Coulomb flux tube vanishes exponentially with the transverse distance, and has a width larger than the minimal flux tube. This is difficult to reconcile with theoretical prediction. We performed a somewhat different analysis of the same quantity  by emphasizing large-$y$ values. 
Furthermore by considering other values of $\beta$ we showed that with increasing $\beta$ the genuine, it is likely that Coulomb flux tube with power-law fall-off develops. As a consequence we were concluded that
it  is possible that the energy density profile evolves from the Wilson-like to the Coulomb one in the continuum limit. This was supported by our study of the Euclidean time development of the profile, however, quantitative analysis requires better statistics and in larger volumes.

The power-law  model was not able to describe the small-$y$ data properly, predicting too small values of the energy density near the $Q\bar{Q}$ axis. The same happened with the theoretical prediction of Ref. \citep{Bowman:2004}. This might indicate that on the axis the Coulomb energy density contains a significant non-perturbative contribution which should be explained.


\section{Acknowledgements}

We would like to thank K. Chung and J. Greensite for the helpful correspondence, and for sharing their lattice code with us, and Z. Bryant for useful conversations. This work was supported by the U.S. Department of Energy under Grants No. DE-AC05-06OR23177 and No. DE-FG02-87ER40365. This research was performed on the Indiana University Carbonate supercomputer and was supported in part by Lilly Endowment, Inc., through its support for the Indiana University Pervasive Technology Institute.

\appendix
\newpage 

\section{Plots of the flux tube transverse profile for different $\beta$'s}
\label{app:A}

\begin{figure}[t]
\begin{center}
\subfigure[~$R=1$]
{
\includegraphics[width=0.4\textwidth]{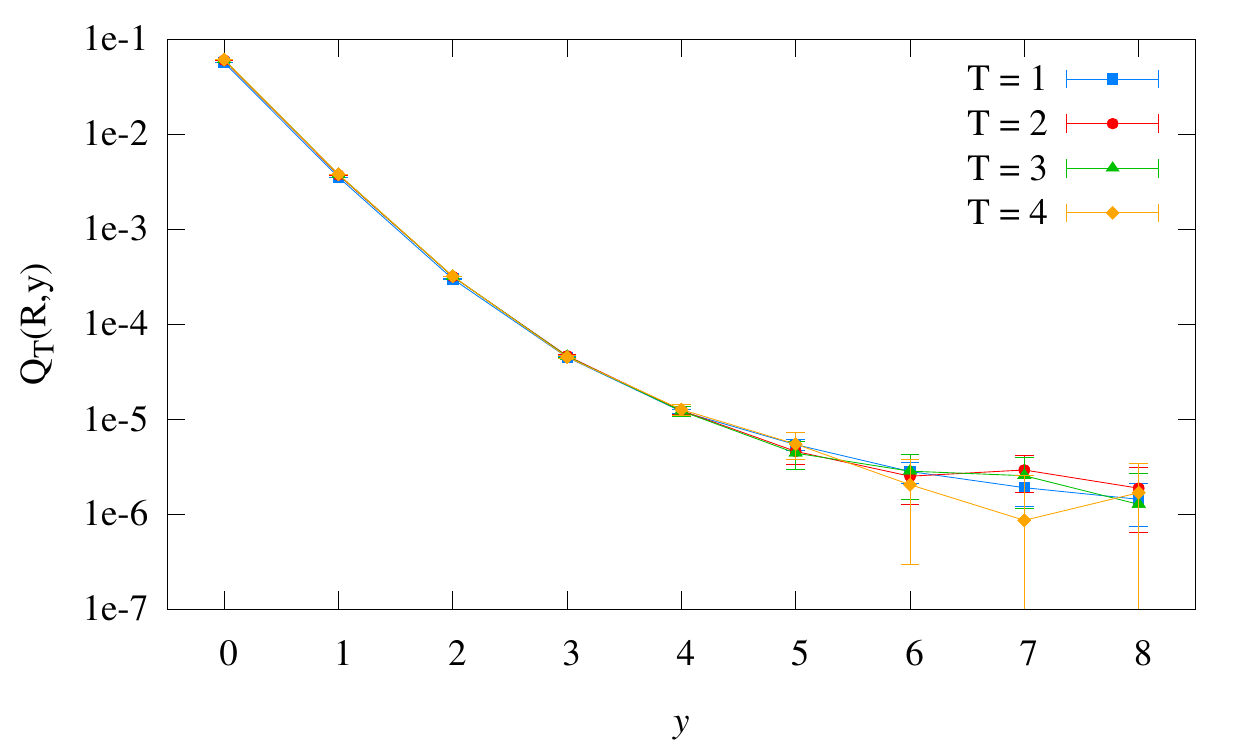}
 
}
\subfigure[~$R=2$]
{
\includegraphics[width=0.4\textwidth]{figures/flux2time.pdf}
 
}
\subfigure[~$R=3$]
{
\includegraphics[width=0.4\textwidth]{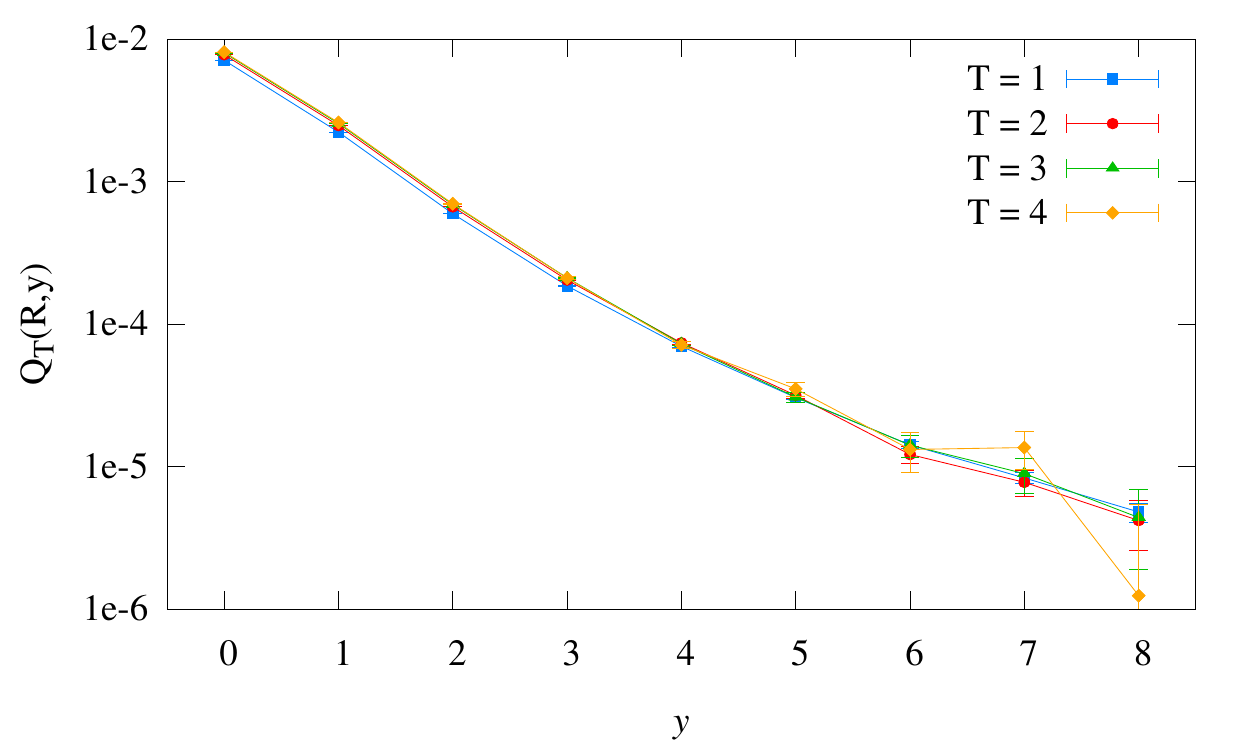}
 
}
\subfigure[~$R=4$]
{
\includegraphics[width=0.4\textwidth]{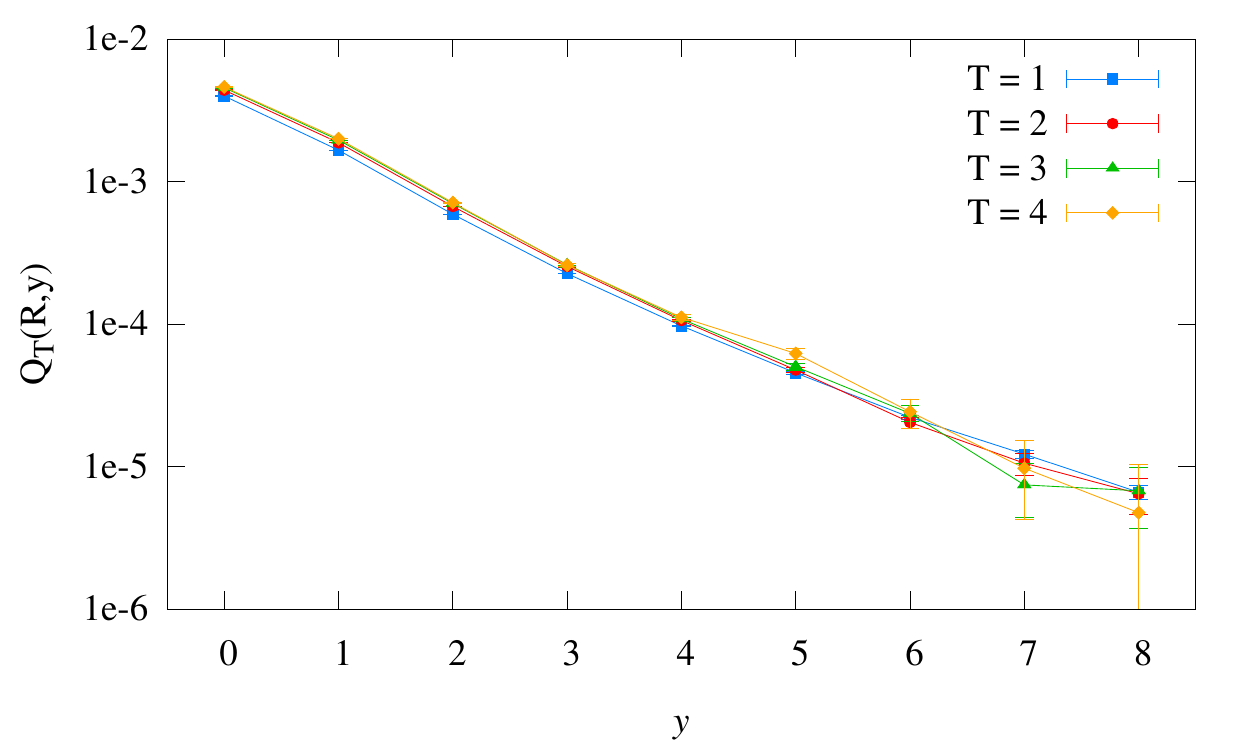}
 
}
\subfigure[~$R=5$]
{
\includegraphics[width=0.4\textwidth]{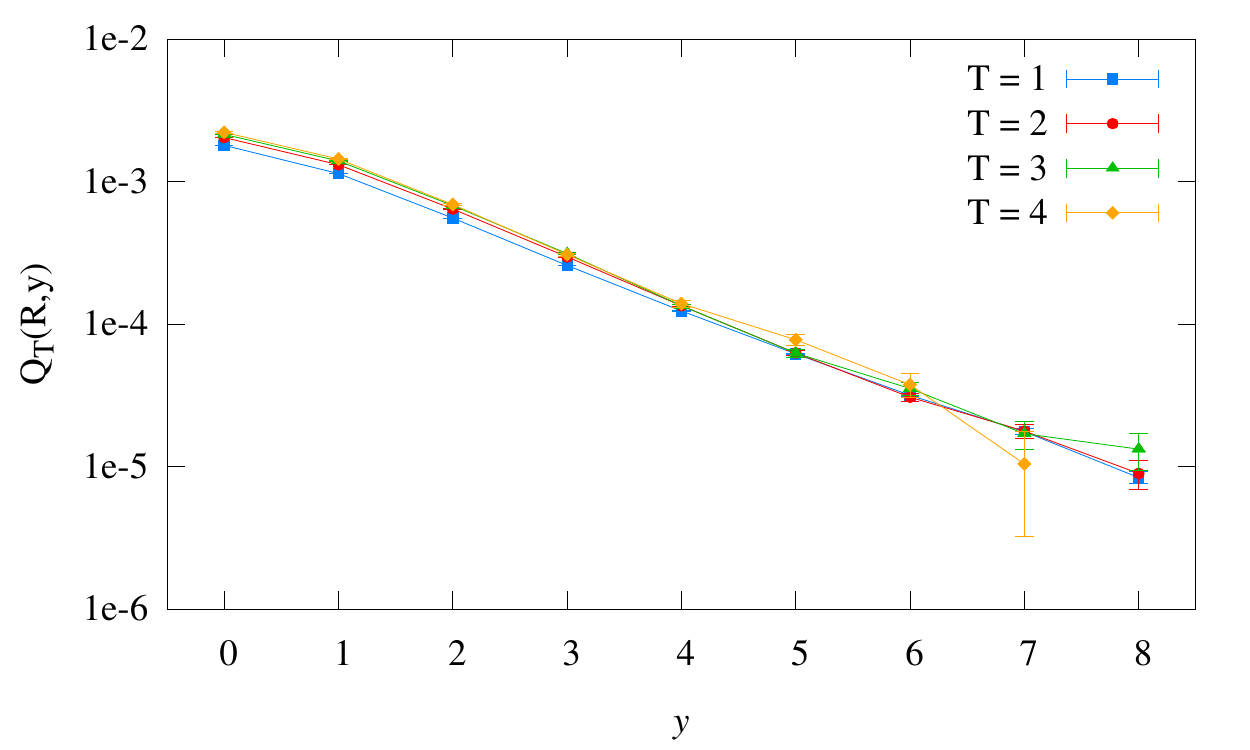}
 
}
\subfigure[~$R=6$]
{
\includegraphics[width=0.4\textwidth]{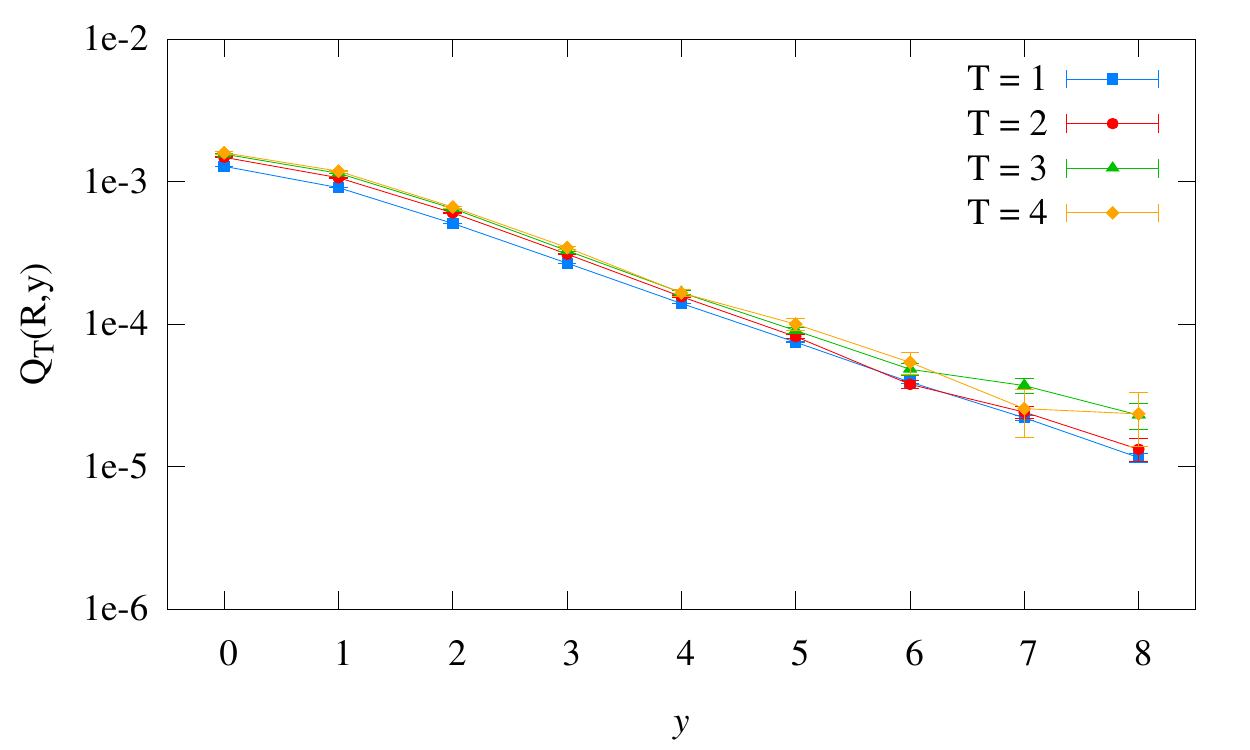}
 
}
\subfigure[~$R=7$]
{
\includegraphics[width=0.4\textwidth]{figures/flux7time.pdf}
 
}
\subfigure[~$R=8$]
{
\includegraphics[width=0.4\textwidth]{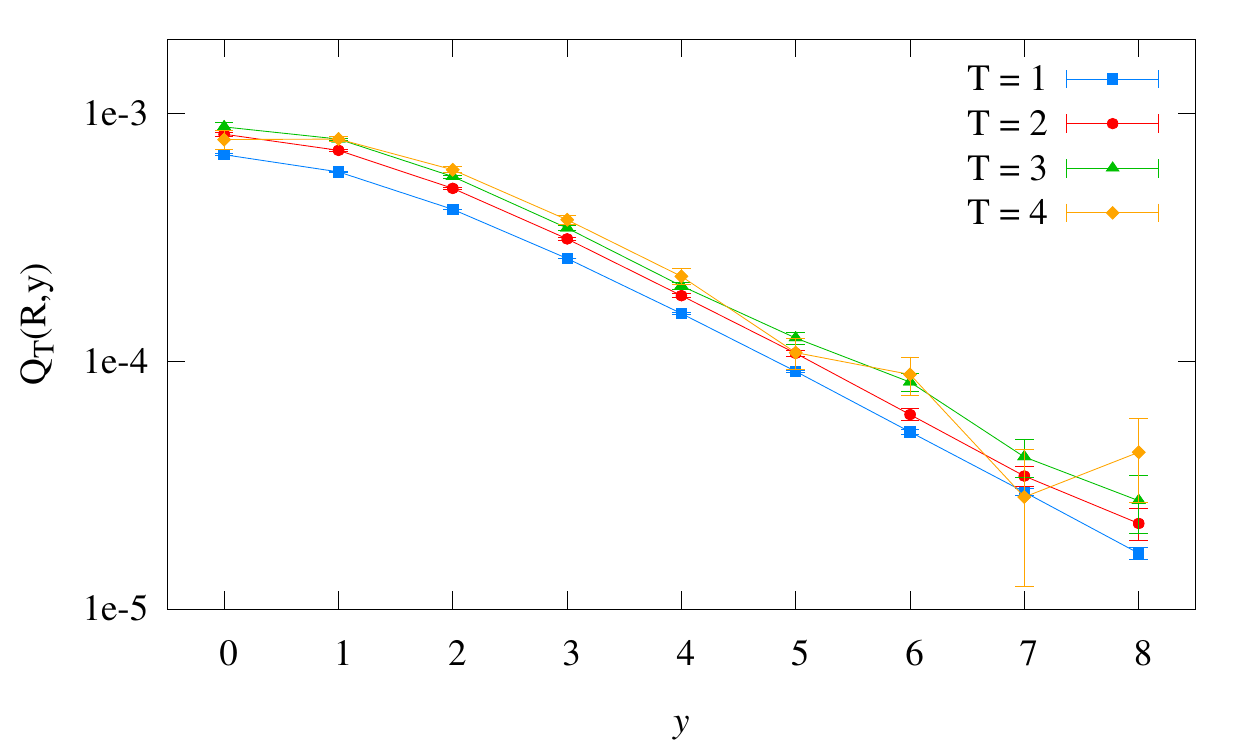}
 
}
\end{center}
\caption{Results for $Q_{T}(R,y)$ for different quark sperations $R$ and different $T$'s, obtained for $\beta=2.5$ and around $11\hspace{2pt} 000$ lattice configurations. Lines joining the data points are included to guide the eye. The data point for $R=5$, $y=8$ and $T=4$ was negative, so is not visible on the semi-log scale. }
\label{fig:all-times}
\end{figure}

On the next pages, we present our data set of $Q_{T=1}(R,y)$ for all values of $\beta$, for $R, y \in [1,8]$, together with the fits. Also, in Fig. \ref{fig:all-times} we show the plots of $Q_{T}(R,y)$ at $\beta=2.5$ for different times $T$.

Apart from the CG model of Eq. \eqref{eq:CG} at $\beta=2.5$, the fits for each $R$ were performed in an interval $y\in[y_\text{min},8]$, with $y_\text{min}=1, 2$ or $3$ depending on $R$.  We claim it is important to include large values of $y$ (as long as they are not affected by the finite volume effects) to discriminate between power-law and exponential behavior. In general only for large transverse distances $y$ the difference between various exponential and power-law models might become significant. Moreover, examining Fig. \ref{fig:5R6}, one can see that for small $y$ and large $R$ the lines have curvature and are flatter compared to a simple exponential. This indicates that close to the quark-antiquark axis the exponential model (CG) may not be accurate. Thus the small $y$ should be excluded when fitting with this model. In practice, none of the models we used was successfully in describing the data $y$-dependence close to the $Q\bar{Q}$ axis, and as a consequence, to obtain satisfactory values of $\chi^2/$d.o.f. We had to remove points $y=0,1$, and sometimes $y=2$, from fit intervals. These points usually come with a small relative errors, thus they affect the value of $\chi^2$ significantly. The fit parameters are summarized in Tabs. \ref{tab:CG25}, \ref{tab:PL25}, \ref{tab:23CG}, \ref{tab:PL27}.

As an illustration of how a choice of the fitting interval affects the quality of the fit, for the exponential model $Q_\text{CG}$ we followed the fit procedure of Ref. \citep{Chung:2017}. Specifically, we used the same intervals as in \citep{Chung:2017} for small $R$, up to $R=5$. For larger values of $R$, one may argue that as function of $y$, $Q(R,y)$ falls off exponentially for $y \geqslant 2$ and we excluded the points  $y=0,1$ from the fit. Even with such an ``optimized" data set, we find the  resulting values of  $\chi^2/$d.o.f to be quite large as shown in Tab. \ref{tab:CG25}.


\onecolumngrid

\begin{figure}[h]
\begin{center}
\subfigure[~$R=1$]
{
\includegraphics[width=0.4\textwidth]{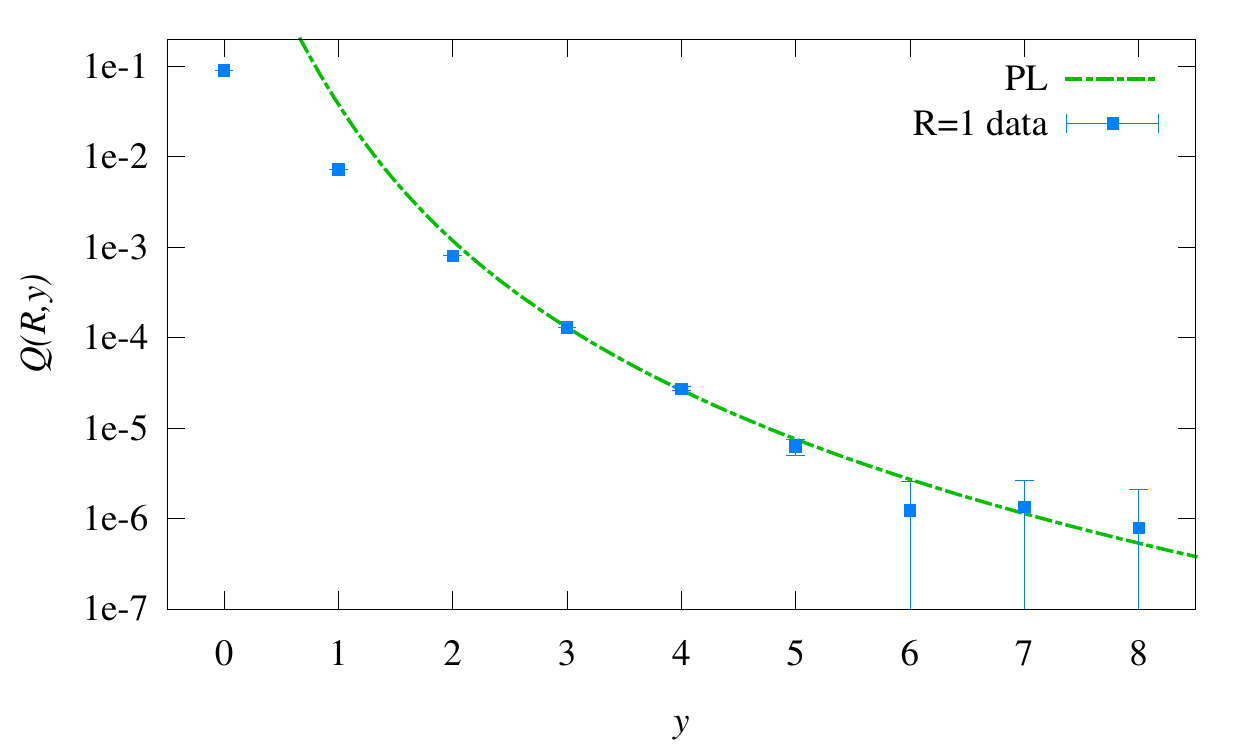}
 
}
\subfigure[~$R=2$]
{
\includegraphics[width=0.4\textwidth]{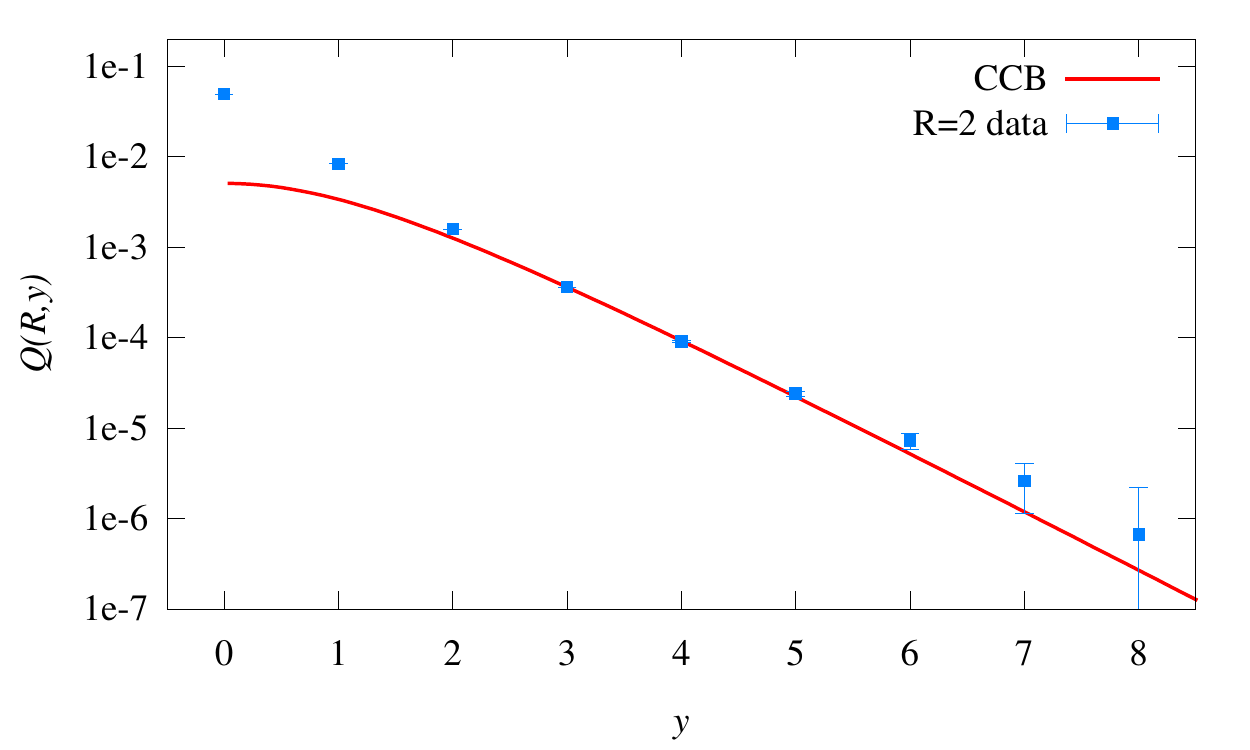}
 
}
\subfigure[~$R=3$]
{
\includegraphics[width=0.4\textwidth]{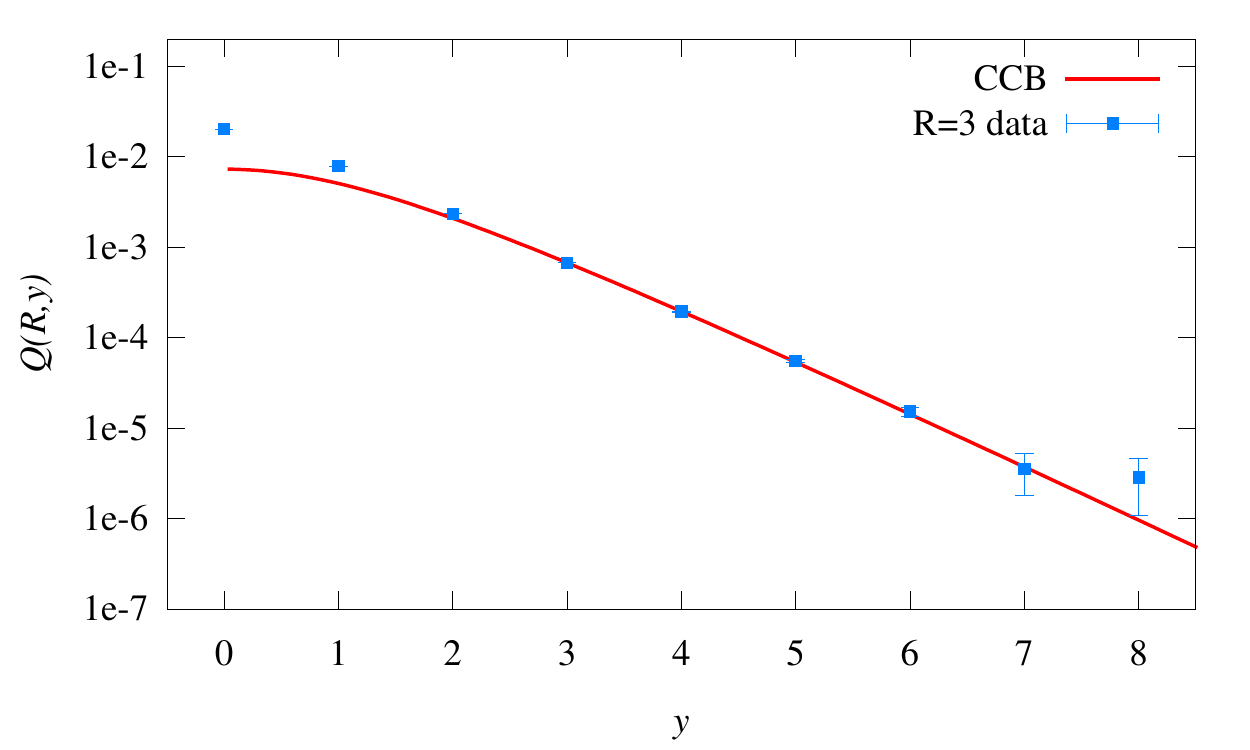}
 
}
\subfigure[~$R=4$]
{
\includegraphics[width=0.4\textwidth]{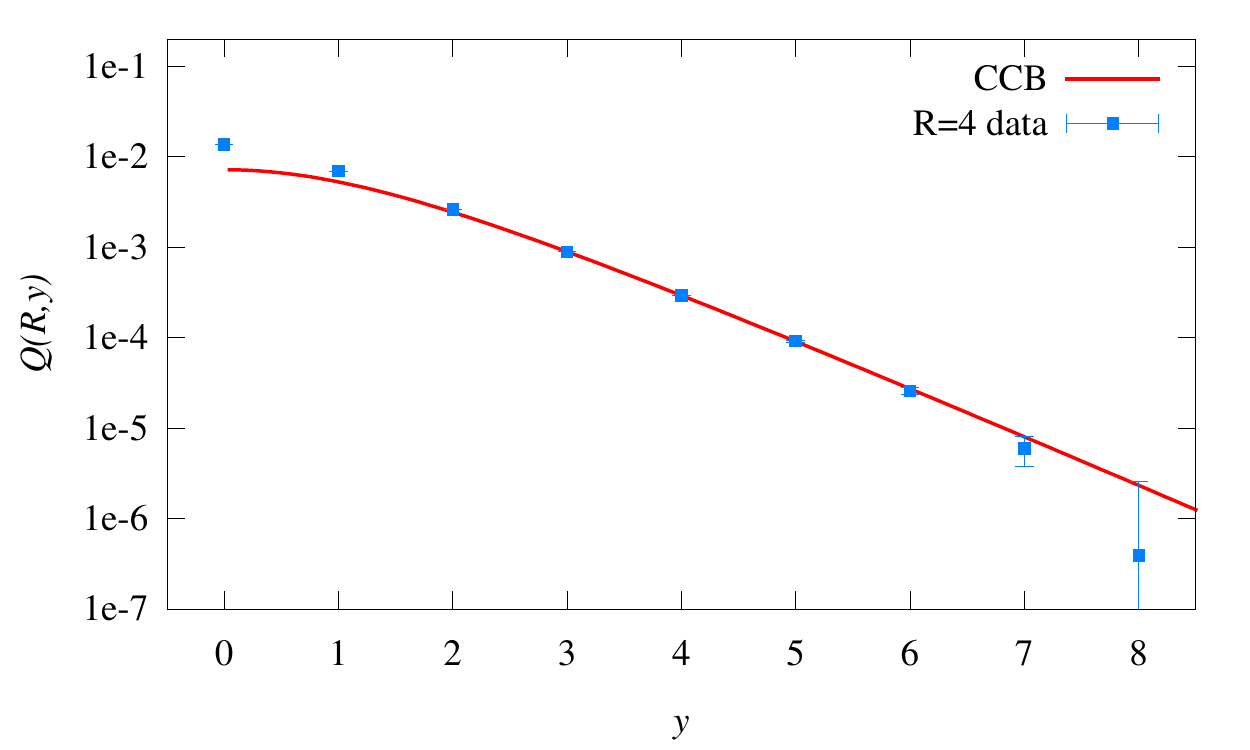}
 
}
\subfigure[~$R=5$]
{
\includegraphics[width=0.4\textwidth]{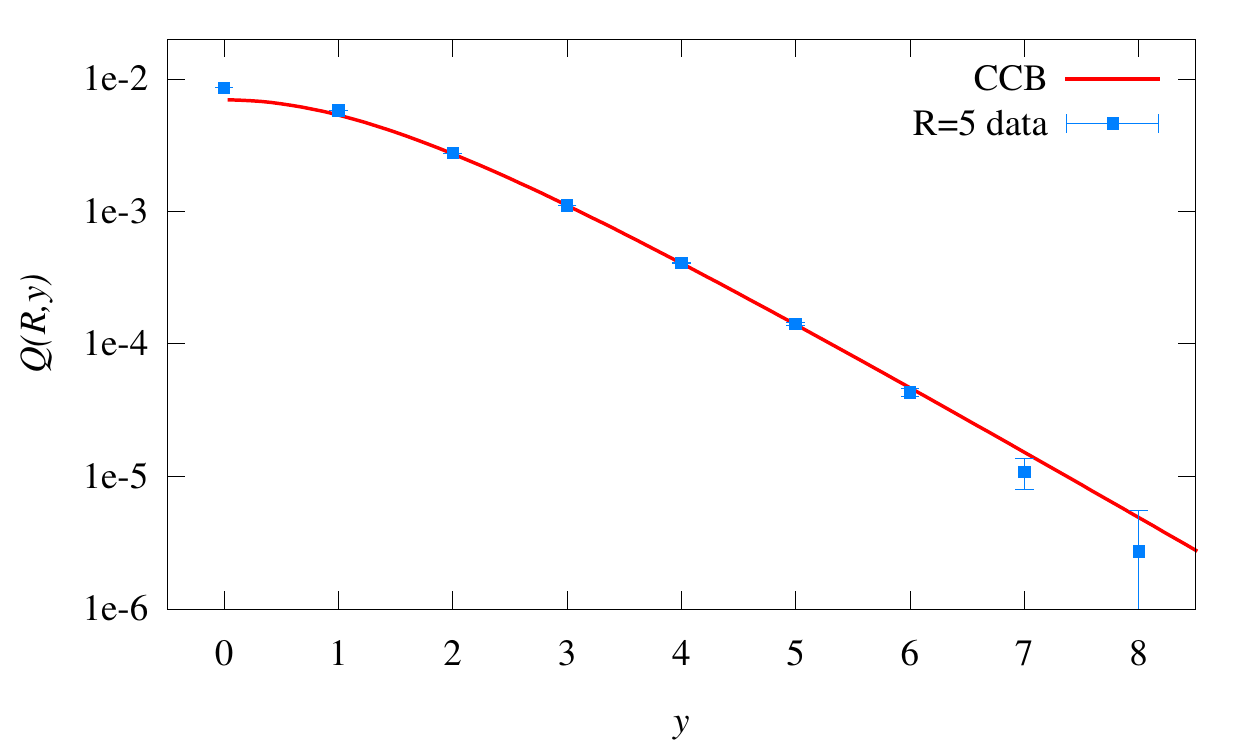}
 
}
\subfigure[~$R=6$]
{
\includegraphics[width=0.4\textwidth]{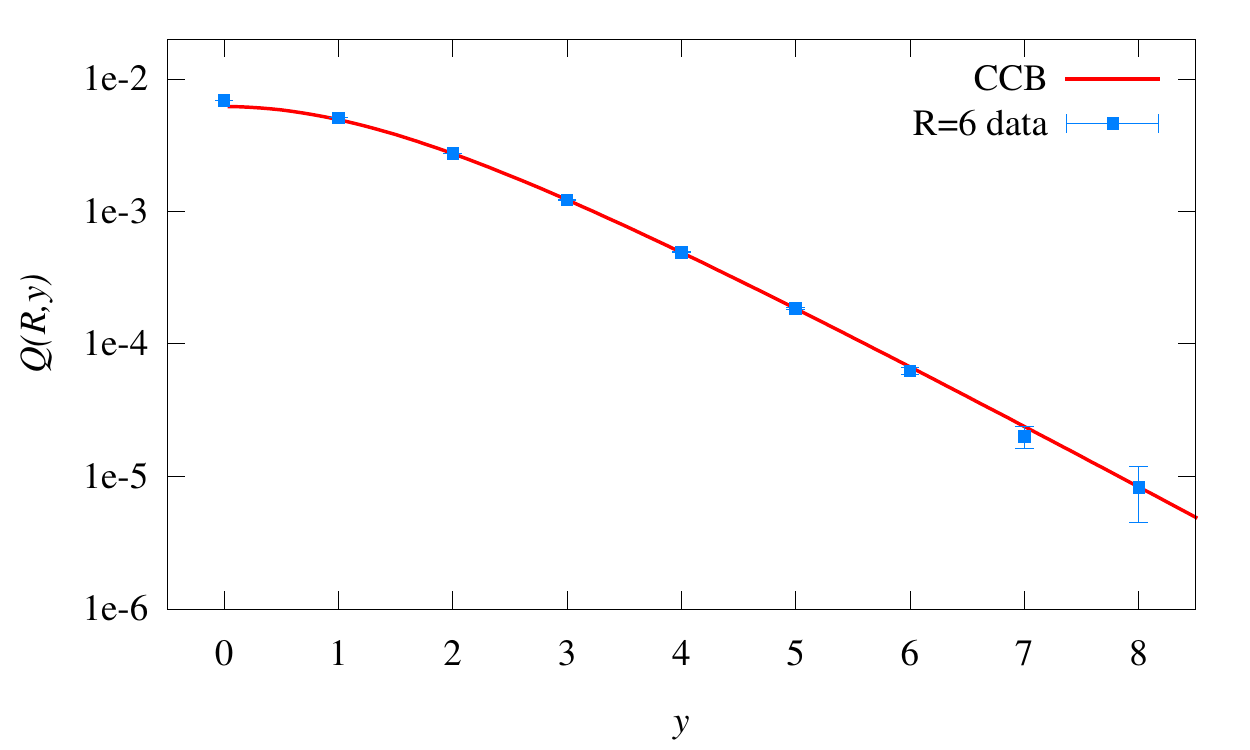}
 
}
\subfigure[~$R=7$]
{
\includegraphics[width=0.4\textwidth]{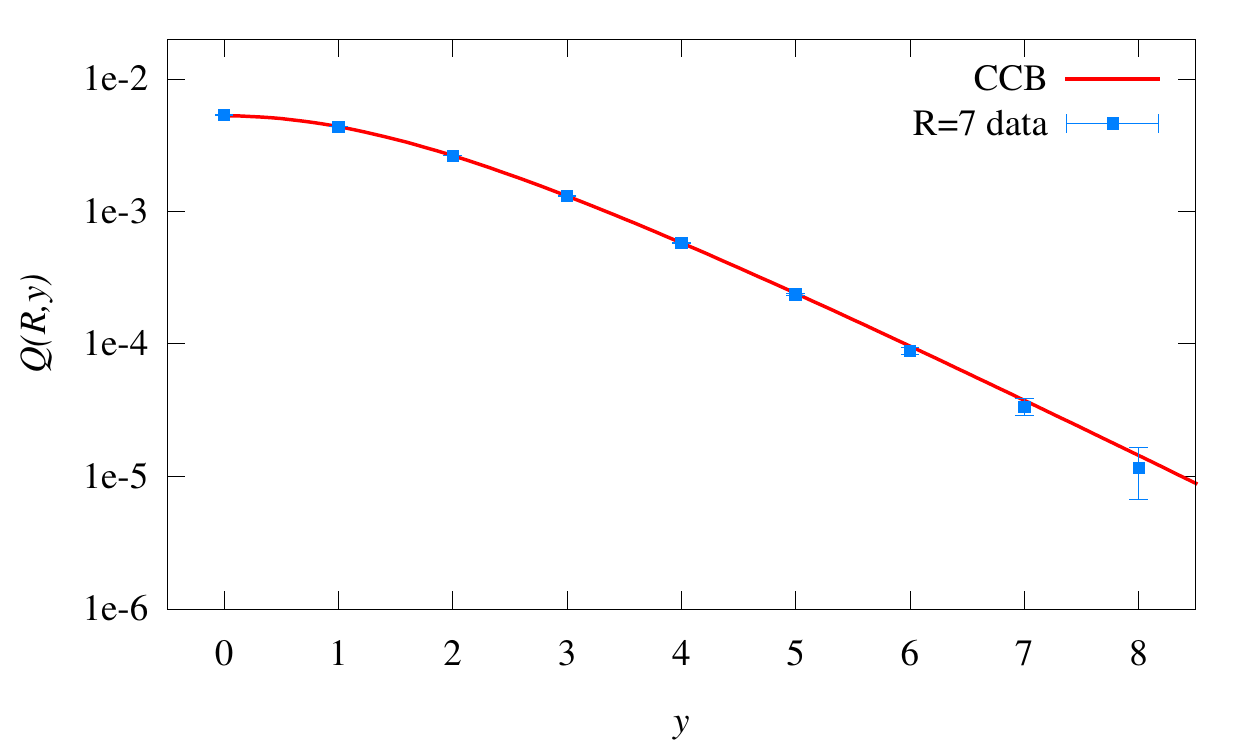}
 
}
\subfigure[~$R=8$]
{
\includegraphics[width=0.4\textwidth]{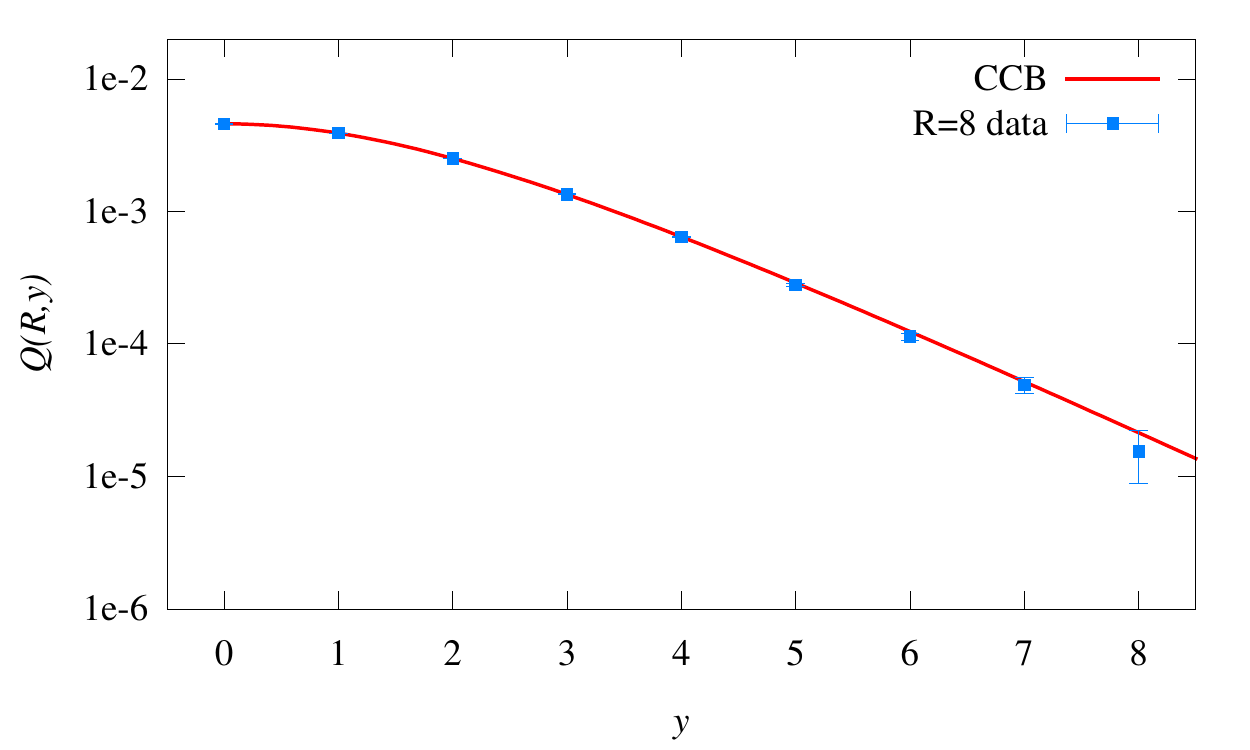}
 
}
\end{center}
\caption{Results for $Q_{T=1}(R,y)$ for different quark separations $R$, obtained for $\beta=2.3$ and 11\hspace{2pt}000 lattice configurations. The red line is the best CCB fit to the data. It can be seen, how the results start agreeing with this model as quark and anti-quark are separated further away. For $R=1$ we used the PL fit in the interval $[3,8]$, obtaining $a=0.23(5)$, $b=2.83(7)$ and $\chi^2/$d.o.f.$=0.71$. The PL model was not able to describe the behavior of the transverse profile for $R \geqslant 2$ and thus is not included on the graphs. The fit parameters are given in Tab. \ref{tab:23CG}}
\label{fig:allR-23}
\end{figure}

\begin{figure}[h]
\begin{center}
\subfigure[~$R=1$]
{
\includegraphics[width=0.4\textwidth]{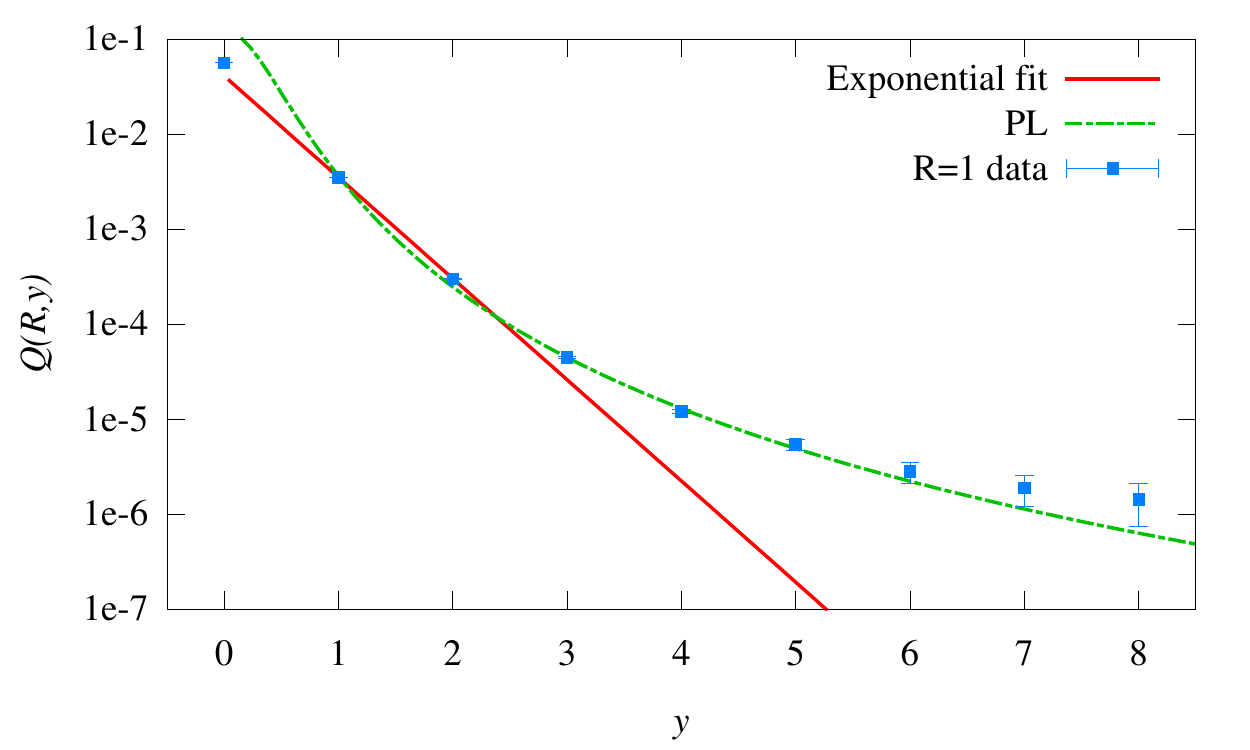}
 
}
\subfigure[~$R=2$]
{
\includegraphics[width=0.4\textwidth]{figures/5R2.pdf}
 
}
\subfigure[~$R=3$]
{
\includegraphics[width=0.4\textwidth]{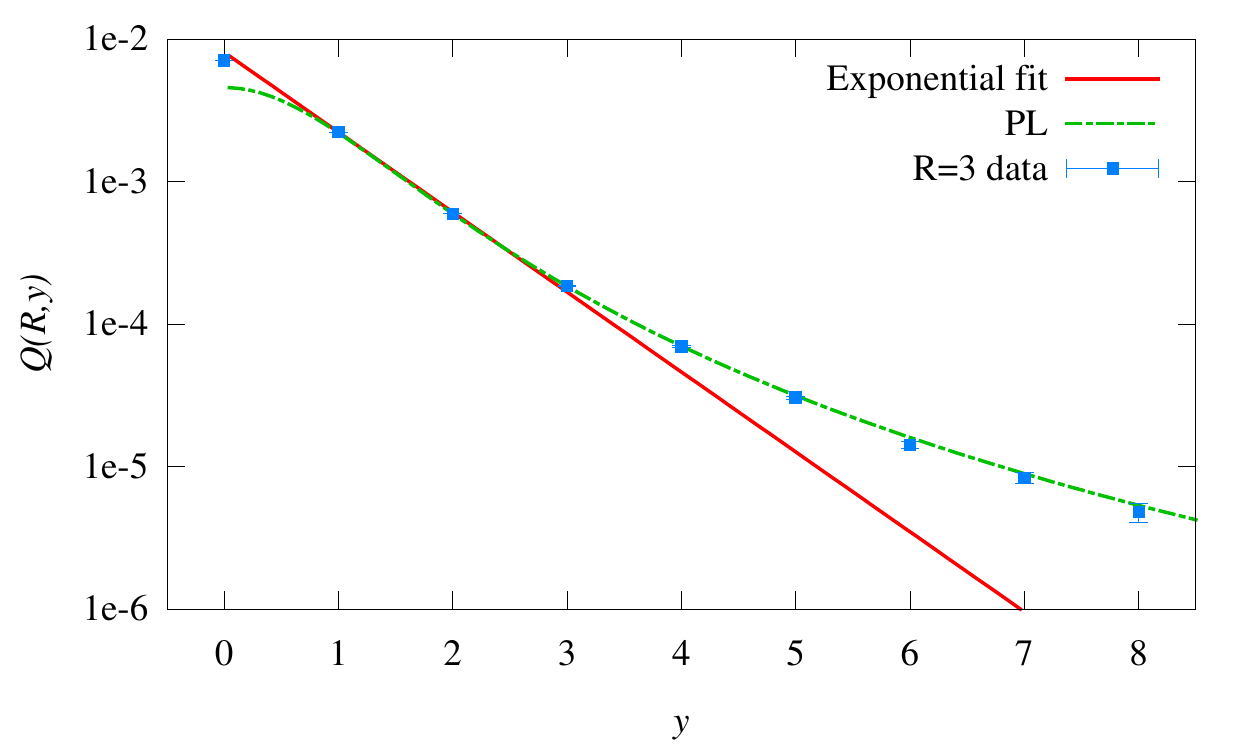}
 
}
\subfigure[~$R=4$]
{
\includegraphics[width=0.4\textwidth]{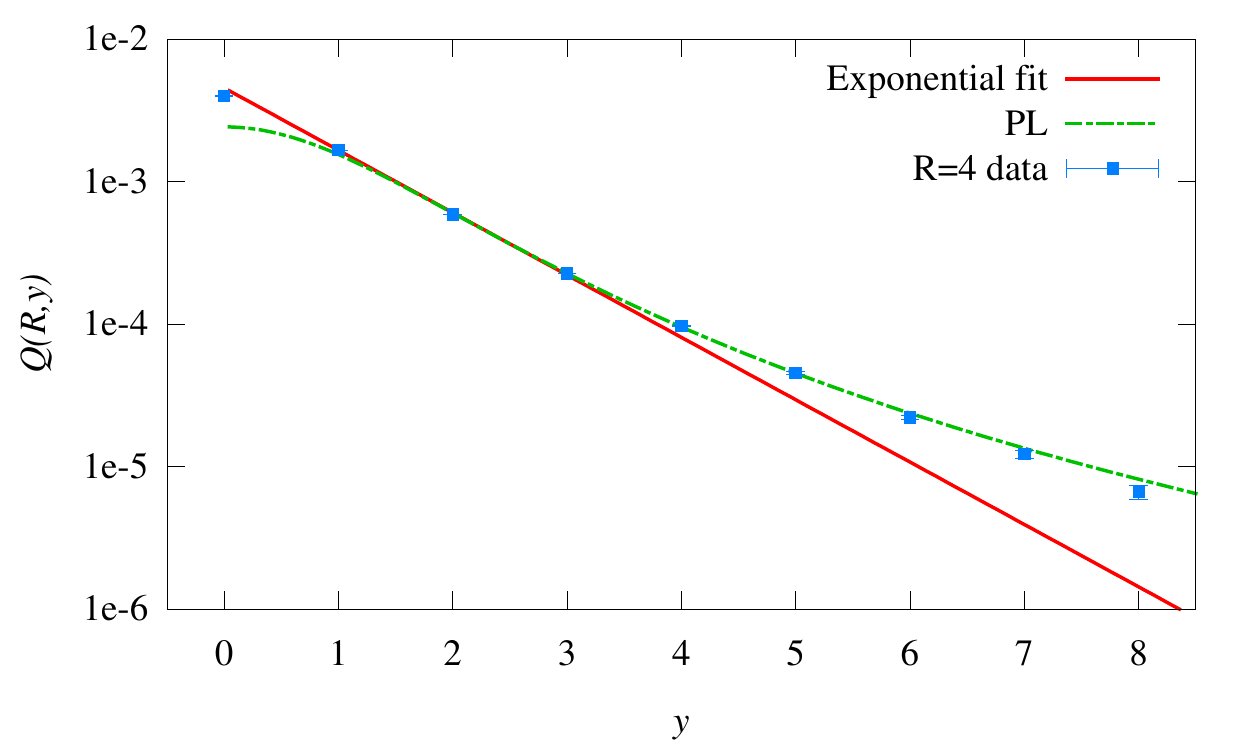}
 
}
\subfigure[~$R=5$]
{
\includegraphics[width=0.4\textwidth]{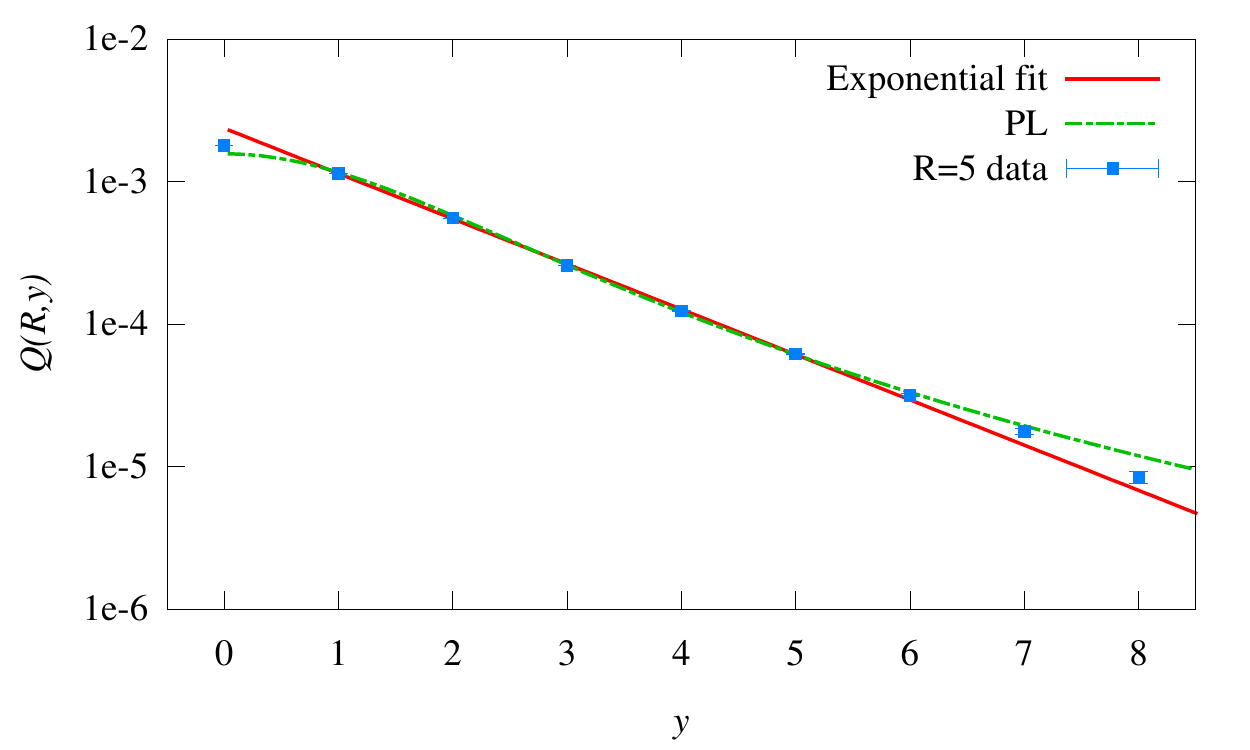}
 
}
\subfigure[~$R=6$]
{
\includegraphics[width=0.4\textwidth]{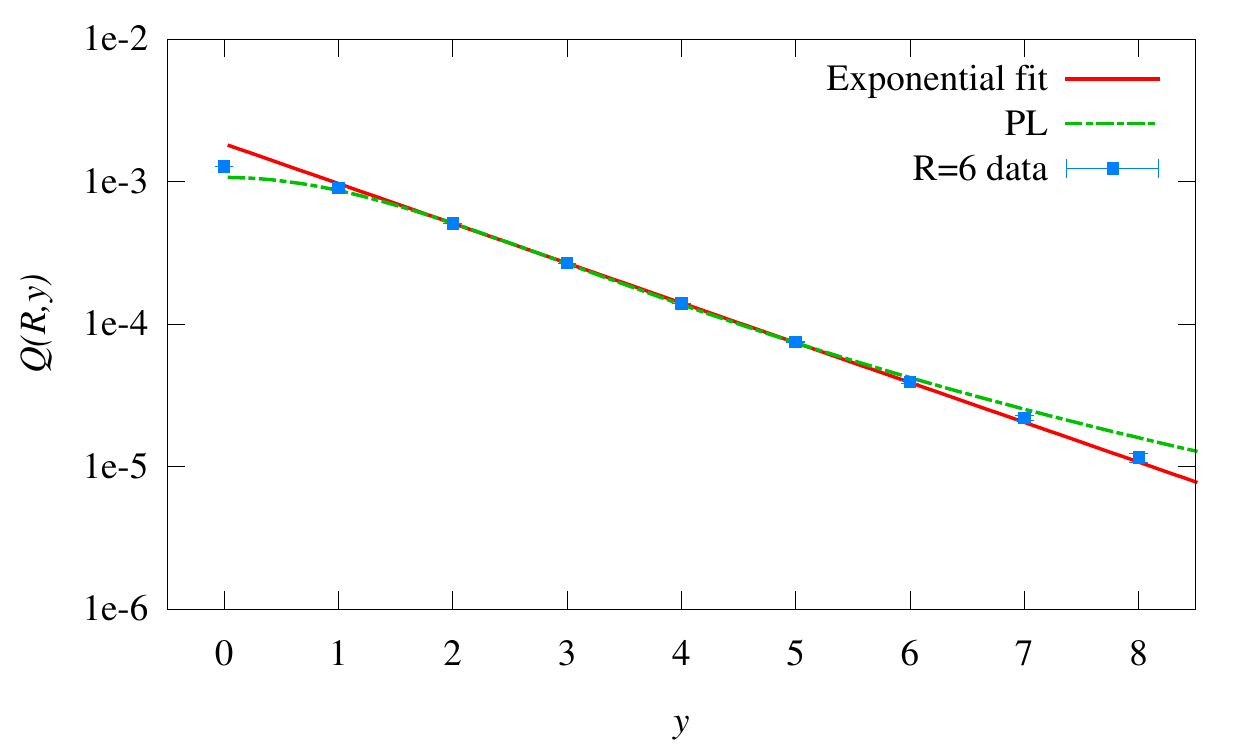}
 
}
\subfigure[~$R=7$]
{
\includegraphics[width=0.4\textwidth]{figures/5R7.pdf}
 
}
\subfigure[~$R=8$]
{
\includegraphics[width=0.4\textwidth]{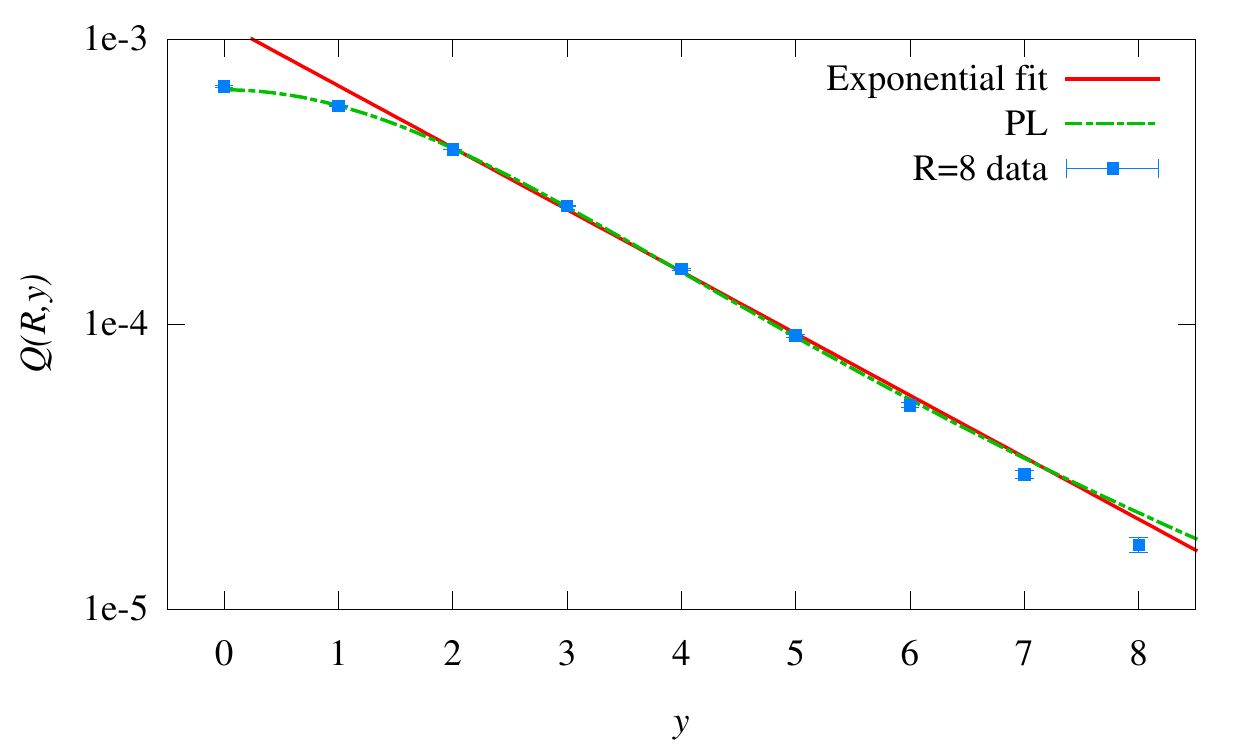}
 
}
\end{center}
\caption{Results for $Q_1(R,y)$ for different quark sperations $R$, obtained for $\beta=2.5$ and 30\hspace{2pt}000 lattice configurations. The red, solid line is the best exponential (CG) fit to the data, performed following the procedure of Ref. \citep{Chung:2017} for $R \leqslant 5$. The green, dashed line is the best PL fit. Fit parameters for both fits are given in Tabs. \ref{tab:CG25} and \ref{tab:PL25}.}
\label{fig:allR-25}
\end{figure}

\begin{figure}[h]
\begin{center}
\subfigure[~$R=1$]
{
\includegraphics[width=0.4\textwidth]{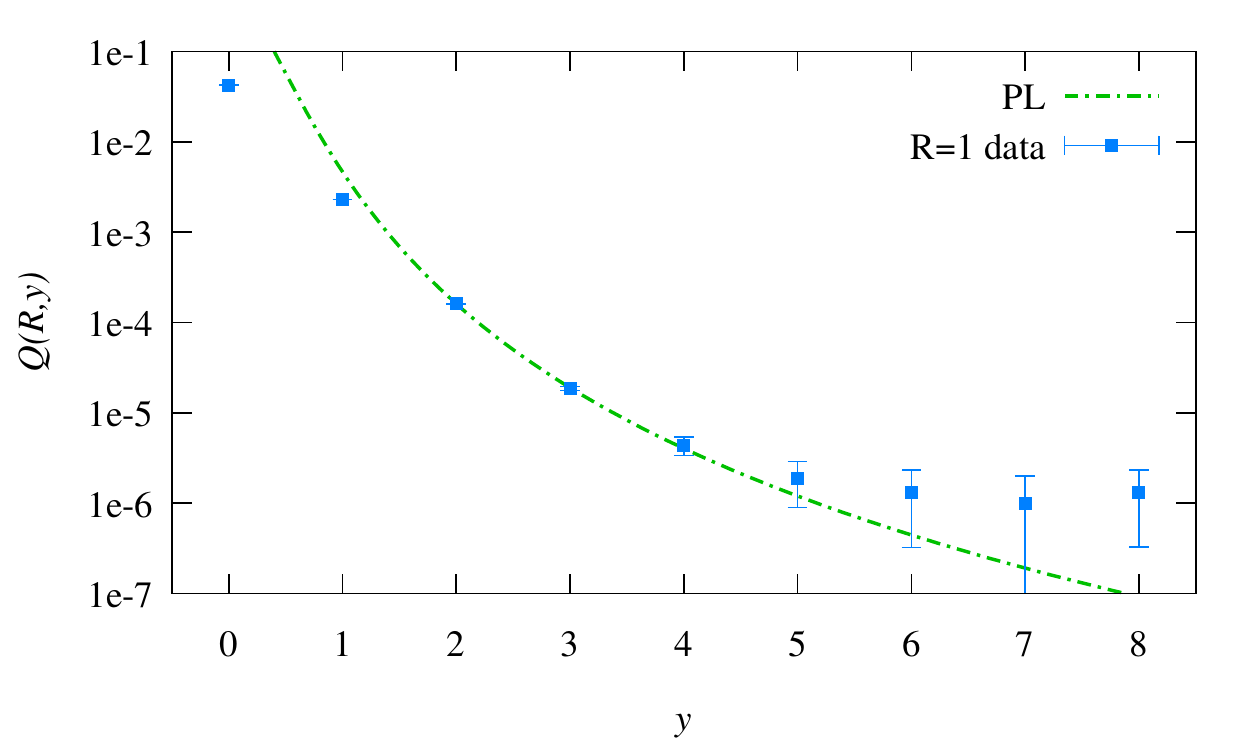}
 
}
\subfigure[~$R=2$]
{
\includegraphics[width=0.4\textwidth]{figures/7RR2.pdf}
 
}
\subfigure[~$R=3$]
{
\includegraphics[width=0.4\textwidth]{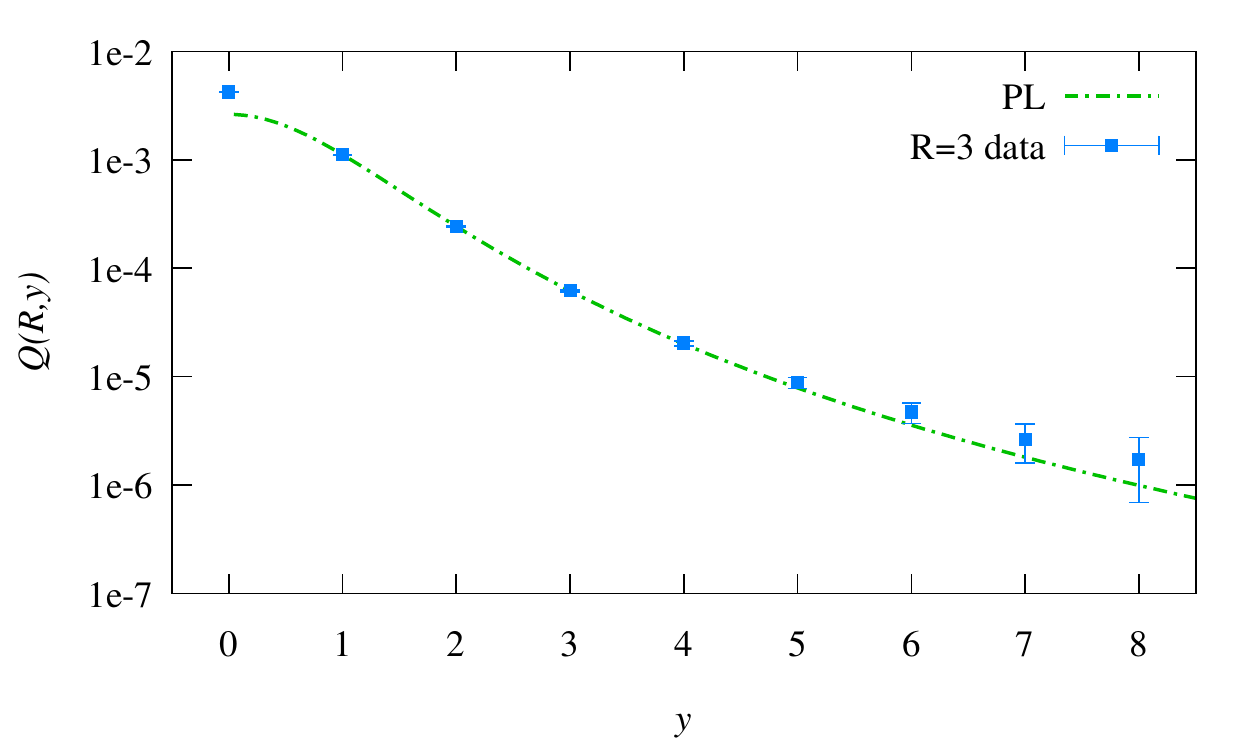}
 
}
\subfigure[~$R=4$]
{
\includegraphics[width=0.4\textwidth]{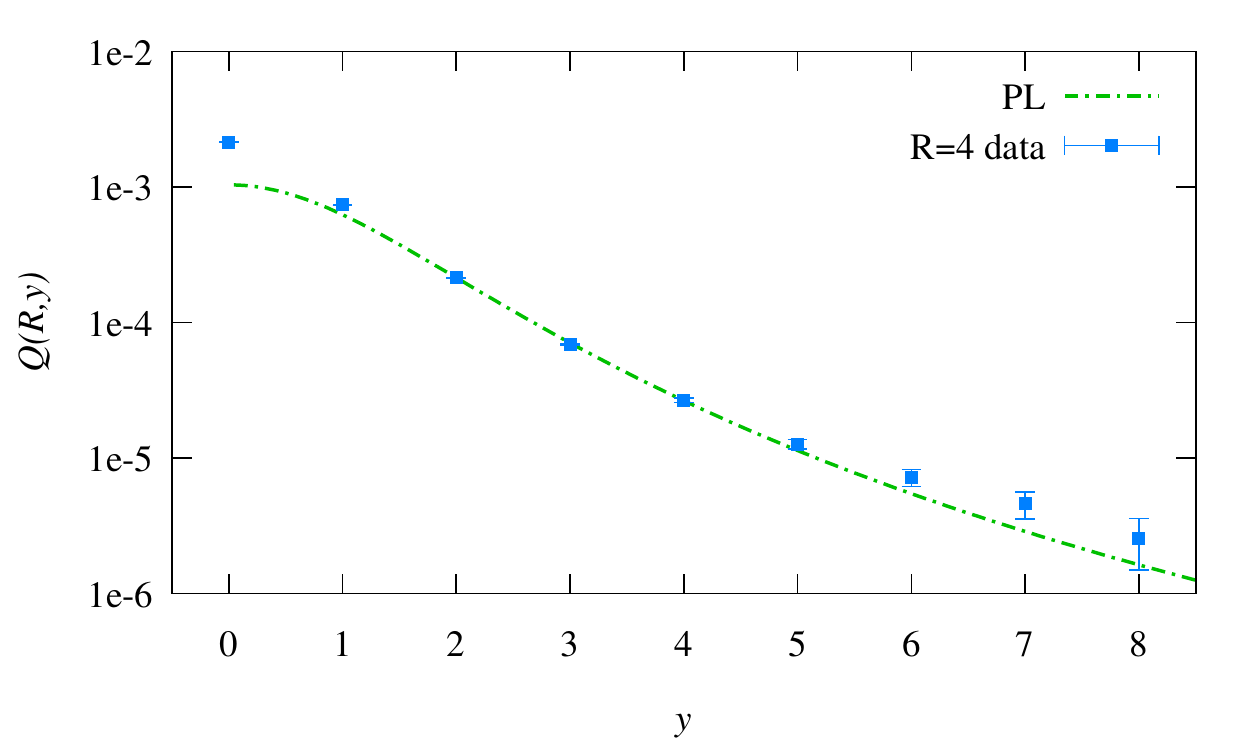}
 
}
\subfigure[~$R=5$]
{
\includegraphics[width=0.4\textwidth]{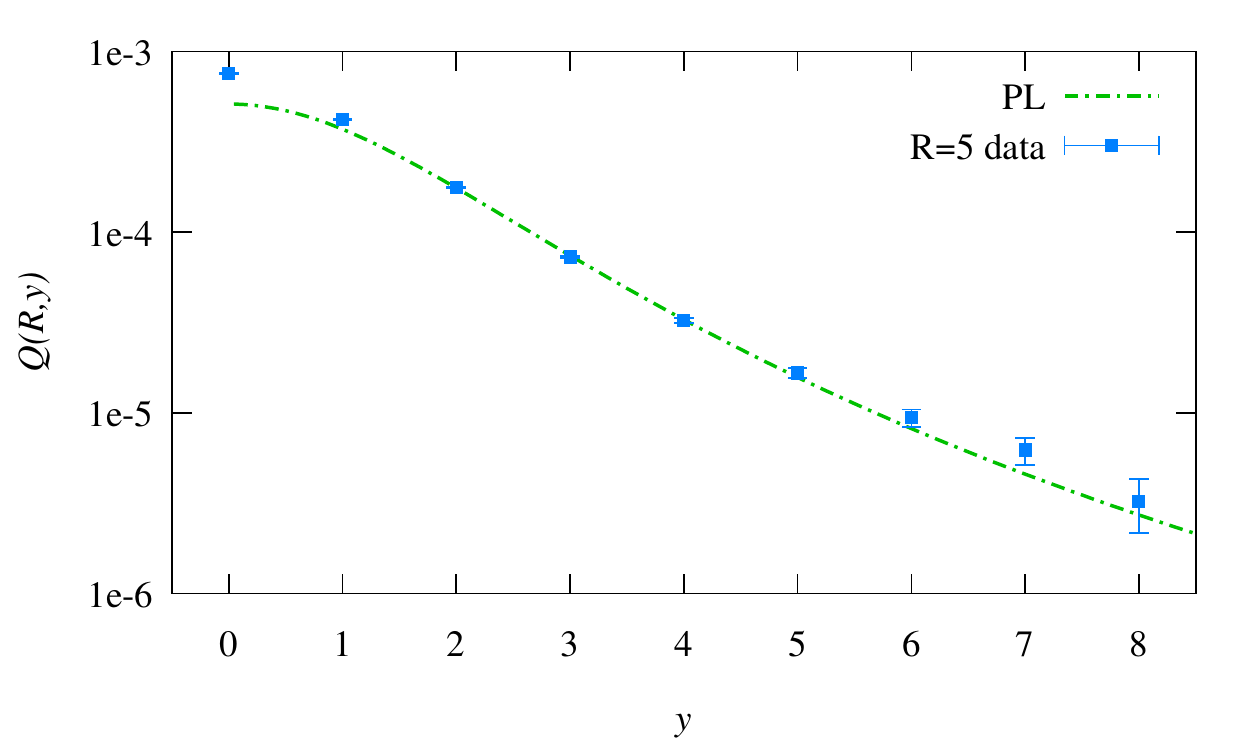}
 
}
\subfigure[~$R=6$]
{
\includegraphics[width=0.4\textwidth]{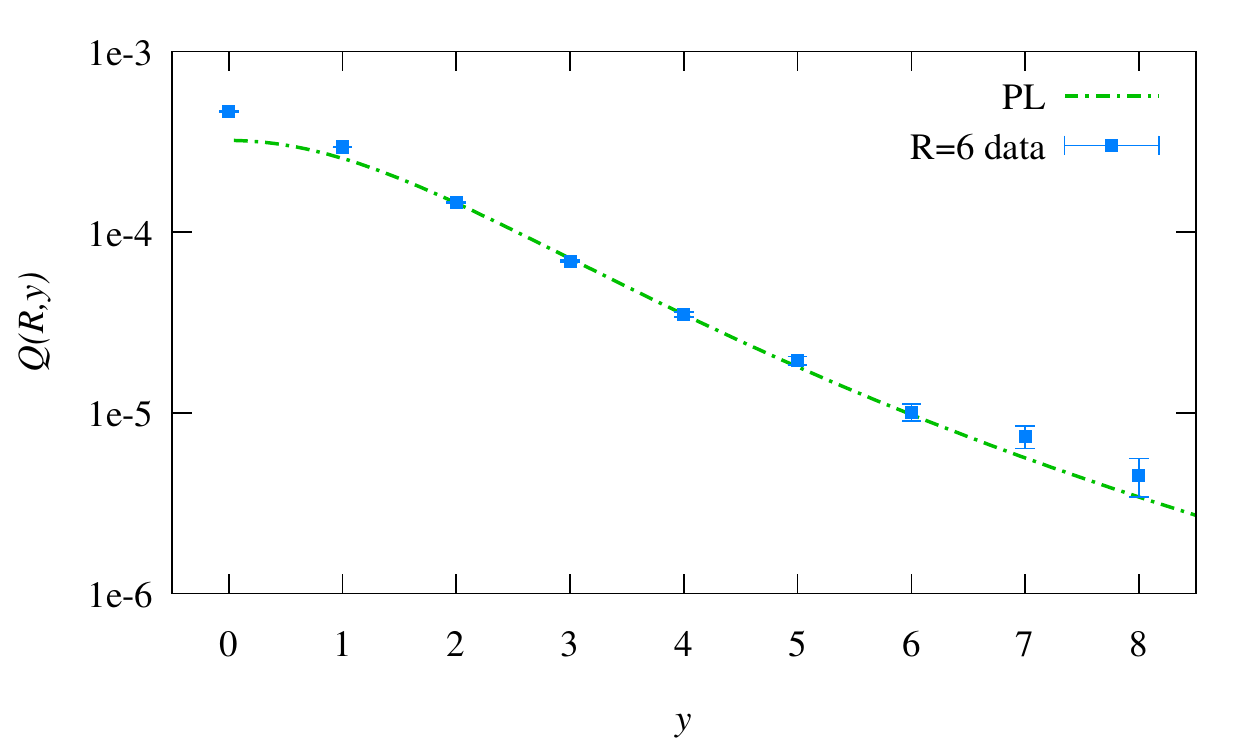}
 
}
\subfigure[~$R=7$]
{
\includegraphics[width=0.4\textwidth]{figures/7RR7.pdf}
 
}
\subfigure[~$R=8$]
{
\includegraphics[width=0.4\textwidth]{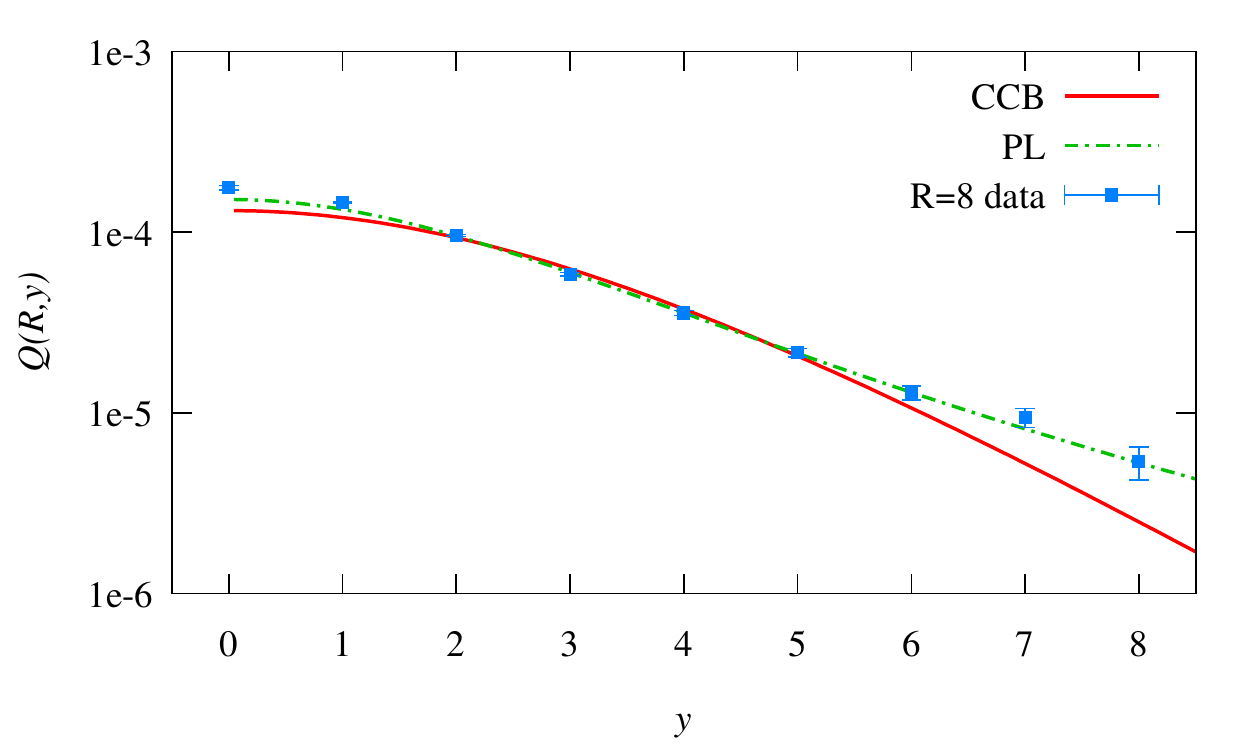}
 
}
\end{center}
\caption{Results for $Q_1(R,y)$ for different quark sperations $R$, obtained for $\beta=2.7$ and around 11\hspace{2pt}000 lattice configurations. The green, dashed line is the best PL fit. We also present the CCB fit (red, solid line) for $R=7$ and $R=8$, performed in the interval $y\in[2,8]$ and yielding $\chi^2/$d.o.f. equal to 14.53 and 8.73, respectively. The fit parameters for the PL fit are given in Tab. \ref{tab:PL27}.}
\label{fig:allR-27}
\end{figure}

\twocolumngrid
\clearpage

\bibliographystyle{apsrev4-1}
\bibliography{coulomb}

\end{document}